\documentclass[11pt]{article}
\usepackage[usenames,dvipsnames,svgnames,table]{xcolor} 
\usepackage[obeyspaces,hyphens,spaces]{url}
\usepackage{jcapmod}
\usepackage[english]{babel}
\usepackage{amsmath, amssymb, amsbsy, amstext, amsthm,mathtools,verbatim}
\usepackage{hyperref}
\usepackage{graphicx}
\usepackage{amsfonts}
\usepackage{fouridx}
\usepackage{cases}
\usepackage{bm}
\usepackage{bbm}
\usepackage{cleveref}
\usepackage{slashed}
\usepackage{mdframed}
\usepackage{tensor}

\usepackage{setspace}

\usepackage{braket}

\usepackage{subcaption}

\usepackage{tikz}
\usetikzlibrary{calc,shadings,patterns,tikzmark,fadings, decorations.markings, decorations.pathmorphing}

\usepackage{mathrsfs}

\definecolor{Blue}{rgb}{0.25, 0.41, 0.88}
\definecolor{Red}{rgb}{0.92,0.,0.}
\definecolor{darkorange}{rgb}{1.0,0.549,0.}
\definecolor{cobalt}{RGB}{44, 98, 120}
\definecolor{Mathematica1}{rgb}{0.368417, 0.506779, 0.709798}
\definecolor{Mathematica2}{rgb}{0.880722, 0.611041, 0.142051}
\definecolor{Mathematica3}{rgb}{0.560181, 0.691569, 0.194885}
\definecolor{Mathematica4}{rgb}{0.922526, 0.385626, 0.209179}
\definecolor{Mathematica5}{rgb}{0.528488, 0.470624, 0.701351}
\definecolor{Mathematica6}{rgb}{0.772079, 0.431554, 0.102387}
\definecolor{Mathematica7}{rgb}{0.363898, 0.618501, 0.782349}
\definecolor{Mathematica8}{rgb}{1, 0.75, 0}
\definecolor{Mathematica9}{rgb}{0.647624, 0.37816, 0.614037}
\definecolor{plotBlue}{RGB}{94, 130, 181}
\definecolor{plotRed}{RGB}{233, 85, 54}
\definecolor{plotGreen}{RGB}{142, 176, 50}
\definecolor{plotPurple}{RGB}{135, 120, 178}

\newcolumntype{C}[1]{>{\centering\let\newline\\\arraybackslash\hspace{0pt}}m{#1}}

\def\e{{\lab{e}}}

\makeatletter
\newlength{\apb@width}
\newcommand{\autoparbox}[2][c]{\settowidth{\apb@width}{#2}\parbox[#1]{\apb@width}{#2}}

\makeatother

\makeatletter
\newsavebox\myboxA
\newsavebox\myboxB
\newlength\mylenA

\newcommand*\xoverline[2][0.75]{
    \sbox{\myboxA}{$\m@th#2$}%
    \setbox\myboxB\null
    \ht\myboxB=\ht\myboxA%
    \dp\myboxB=\dp\myboxA%
    \wd\myboxB=#1\wd\myboxA
    \sbox\myboxB{$\m@th\overline{\copy\myboxB}$}
    \setlength\mylenA{\the\wd\myboxA}
    \addtolength\mylenA{-\the\wd\myboxB}%
    \ifdim\wd\myboxB<\wd\myboxA%
       \rlap{\hskip 0.5\mylenA\usebox\myboxB}{\usebox\myboxA}%
    \else
        \hskip -0.5\mylenA\rlap{\usebox\myboxA}{\hskip 0.5\mylenA\usebox\myboxB}%
    \fi}
\makeatother

\numberwithin{equation}{section}

\def\beq{\begin{equation}}
\def\eeq{\end{equation}}

\def\bea{\begin{eqnarray}}
\def\eea{\end{eqnarray}}

\def\dd{{\rm d}}
\def\Tr{{\rm Tr}}

\def\beq{\begin{equation}}
\def\eeq{\end{equation}}
\def\bea{\begin{eqnarray}}
\def\eea{\end{eqnarray}}

\def\dd{{\rm d}}
\def\Tr{{\rm Tr}}

\newcommand{\ud}{\mathrm{d}}
\newcommand{\lab}[1]{{\mathrm{#1}}}
\newcommand{\mb}[1]{{\mathbf{#1}}}
\newcommand{\minus}{{\scalebox {0.75}[1.0]{$-$}}}
\newcommand{\sminus}{{\scalebox {0.5}[0.85]{$-$}}}
\newcommand{\spm}{{\scalebox {0.75}[0.75]{$\pm$}}}

\theoremstyle{definition}

\DeclareRobustCommand{\SkipTocEntry}[4]{}

\newcommand{\es}{\hspace{0.5pt}}

\definecolor{blue2}{cmyk}{1, 0.1, 0.1, 0.1}
\definecolor{byzantium}{rgb}{0.44, 0.16, 0.39}

\definecolor{pyBlue}{RGB}{31, 119, 180}
\definecolor{pyRed}{RGB}{214, 39, 40}
\definecolor{pyGreen}{RGB}{44, 160, 44}
\definecolor{pyBlue2}{RGB}{0, 111, 237}
\definecolor{pyRed2}{RGB}{224, 52, 36}

\newcommand{\slab}[1]{{\textsc{#1}}}

\newcommand{\subp}{{\scriptscriptstyle +}}
\newcommand{\subm}{{\scriptscriptstyle -}}

\begin{document}

\pagenumbering{roman}
\begin{titlepage}
\baselineskip=15.5pt \thispagestyle{empty}

\bigskip\

\begin{center}
{\fontsize{20}{24}\selectfont  {\bfseries Zero Modes of Massive Fermions \\[8pt] Delocalize from Axion Strings}}
\end{center}
\vspace{0.3cm}
\begin{center}
{\fontsize{12}{18}\selectfont Hengameh Bagherian,  Katherine Fraser, Samuel Homiller, and John Stout} 

\vspace{10pt}

\textit{Department of Physics, Harvard University, Cambridge, MA 02138, USA}

\end{center}

\vspace{1.2cm}
\hrule \vspace{0.3cm}
\noindent {\bf Abstract}\\[0.1cm]
Massless chiral excitations can arise from the interactions between a fermion and an axion string, propagating along the string and allowing it to superconduct. The properties of these excitations, or zero modes, dictate how the string interacts with light and can thus have important phenomenological consequences. In this paper, we add a nowhere-vanishing Dirac mass for the fermion in the usual model of axion electrodynamics. We find that the zero modes exhibit an interesting phase structure in which they delocalize from the string's core as the mass increases, up until a critical value past which they disappear. We study this structure from an analytic perspective, with explicit numerical solutions, and via anomaly inflow arguments. Finally, we derive the two-dimensional effective theory of the zero mode and its interactions with the four-dimensional gauge field and show how this effective theory breaks down as the zero modes delocalize.

\vskip10pt
\hrule
\vskip10pt

\end{titlepage}

\thispagestyle{empty}
\setcounter{page}{2}
\begin{spacing}{1.03}
\tableofcontents
\end{spacing}

\clearpage
\pagenumbering{arabic}
\setcounter{page}{1}

\newpage

\section{Introduction}

Axions---scalar fields with compact field spaces---were originally proposed to solve the strong CP problem~\cite{Peccei:1977ur, Peccei:1977hh, Weinberg:1977ma,Wilczek:1977pj}, but have since become ubiquitous in high-energy physics. 
They have been postulated to serve many roles in beyond the Standard Model physics: 
to be dark matter~\cite{Preskill:1982cy, Dine:1982ah, Abbott:1982af}; 
to be the inflaton that generates exponential expansion in the early universe~\cite{Freese:1990rb}; 
to give rise to matter/antimatter asymmetry~\cite{Alexander:2004us}; and many others.
Axions are also a generic prediction in string theory, where they arise copiously as the zero modes of higher-form fields as a result of the non-trivial topology of the compactified extra dimensions~\cite{Witten:1984dg, Svrcek:2006yi, Arvanitaki:2009fg,Demirtas:2018akl,Demirtas:2021gsq}.

One particularly interesting aspect of axion phenomenology is their connection to topological defects such as monopoles and strings. 
These connections are interesting theoretically, as they can provide mechanisms for generating the axion mass in Abelian gauge theories~\cite{Stout:2020uaf,Fan:2021ntg}.
They can also be important phenomenologically: in post-inflationary axion models, for instance, the axion abundance is determined by radiation from a network of cosmic axion strings~\cite{Vilenkin:1982ks, Sikivie:1982qv, Davis:1986xc, Harari:1987ht, Shellard:1987bv, Davis:1989nj, Hagmann:1990mj, Battye:1993jv, Battye:1994au, Yamaguchi:1998gx, Klaer:2017ond, Gorghetto:2018myk, Vaquero:2018tib, Buschmann:2019icd, Gorghetto:2020qws, Dine:2020pds, Buschmann:2021sdq}.
Other recent work on potential signatures of axion string networks can be found in~\cite{Agrawal:2019lkr, Benabou:2023ghl}.

Recently, there has been renewed interest in the fact that axion strings can be {\em superconducting}: they support charged zero modes localized to the string core, which lead to a current on the string proportional to an applied electric field.
That axion strings are superconducting has been understood for many years, as it is intimately related to the phenomenon of anomaly inflow elucidated by Callan and Harvey~\cite{Callan:1984sa}, which we will review in what follows. 
A great deal of the physics of anomaly inflow and axion strings was worked out in subsequent years: see~\cite{Kaplan:1987kh, Naculich:1987ci, Manohar:1988gv, Harvey:1988in, Harari:1992ea, Blum:1993yd, Harvey:2000yg, Heidenreich:2021yda, Fukuda:2020imw} 
for a collection of early and important works on the subject.
The contemporary interest in axion string superconductivity is in part due to the realization that these strings could interact with a primordial magnetic field, leading to striking signatures due to the formation of vortons~\cite{Fukuda:2020kym,Ibe:2021ctf} or bound states~\cite{Agrawal:2020euj}.

The existence of the charged zero modes is often explained in the context of simple models with a classical PQ symmetry.
In these models, the fermions acquire mass from the vacuum expectation value of the radial mode of the scalar field whose phase is the axion.
In classical string configurations, the vacuum expectation value goes to zero at the core of the string solution and intuition strongly suggests that massless modes will exist in a region localized at the string core.
An analogous argument successfully explains the existence of zero modes on domain walls, as we will~review.

While this simple picture is intuitive, for axion strings, it is clearly incomplete. 
For one, it pays no heed to the crucial fact that the zero modes on axion strings are {\em chiral}---a property which distinguishes them from zero modes on e.g., Witten strings~\cite{Witten:1984eb}. 
Moreover, in more complicated models (for instance, the DFSZ axion~\cite{Dine:1981rt, Zhitnitsky:1980tq}), there can exist string configurations in which the vacuum expectation value of the scalar does not vanish at the string core~\cite{Abe:2020ure}. 
Despite this, there are arguments that these strings can be superconducting as well.

This picture is also related to another puzzle. What happens to the zero mode when this classical PQ symmetry is badly and explicitly broken? For instance, if the fermions have a very large mass $m$, we should have no trouble completely integrating them out and the axion string should be blind to their existence. In this case, there is no anomaly to inflow, and so these zero modes should not exist in the limit $m \to \infty$. In contrast, since they are chiral we still expect them to exist for small but non-zero $m$. So, there must be some critical value of the mass $m$ at which they cease to exist. What happens to the zero modes near this critical mass?

The goal of this paper is to shed some light on these puzzles.
We will do so by studying a simple model of axion electrodynamics in which the classical PQ symmetry is {\em explicitly} broken by a Dirac mass $m$ for the fermion.
We demonstrate explicitly that there exist zero mode solutions to the equations of motion in the axion string background, and numerically solve for their profile.
We find that when the Dirac mass is roughly equal in size to the mass $\mu$ induced by the scalar field which spontaneously breaks the PQ symmetry,
the profile functions change dramatically: the zero mode becomes completely delocalized from the string and onto a semi-infinite wedge.
We also revisit the anomaly inflow story in the presence of the bulk mass term, clarifying the topological origin of the zero modes even in the absence of a ``classical'' PQ symmetry.
Finally, we derive the low-energy two-dimensional effective theory for the zero mode and calculate the leading higher-derivative interactions with the bulk gauge field. As one would expect, we show that this effective theory completely breaks down as $m \to \mu$.  
While this simple model is not relevant phenomenologically, our hope is that this work can be applied to more realistic models, with potential astrophysical or cosmological effects that can be studied in future work.

It is worth emphasizing that these considerations are entirely distinct from situations in which the zero modes localized on cosmic (non-axionic) strings can acquire mass from the pairing of left- and right-moving modes~\cite{Hill:1986ts}, which can be induced e.g., by finite temperature effects~\cite{Weldon:1982bn}. The mass term we introduce is for the ``full'' theory of fermions propagating in four-dimensional spacetime. The zero modes localized to the string remain massless on topological grounds, as in the original example~\cite{Callan:1984sa}. 
The zero modes we discuss are also distinct from those found in the background of ``$Z$-strings''~\cite{Achucarro:1999it, Jackiw:1981ee, Starkman:2000bq, Stojkovic:2000ix, Starkman:2001tc}. These string configurations are not topologically stable and there is no mechanism that protects the zero modes from pairing up and acquiring a mass~\cite{Naculich:1995cb, Liu:1995at}.

\vspace{6pt}
\noindent \textbf{Outline}\,\, The rest of this paper is structured as follows. In \S\ref{sec:massless}, we review the axion string solution in the usual case, with a global PQ symmetry, and solve for the massless zero modes explicitly. 
In \S\ref{sec:massive}, we break the PQ symmetry with a Dirac mass and demonstrate that the zero modes still exist. 
We discuss how this can be understood by analogy to the existence of zero modes on domain walls in $2+1$ dimensions, present numerical results for the profile of these zero modes on the string, and discuss the behavior in the ``critical'' mass case.
In \S\ref{sec:inflow}, we recap the original anomaly inflow story, with appropriate modifications to account for the mass term.
Finally, in \S\ref{sec:eft}, we derive the low-energy theory of the zero modes on the string, and discuss how it is impacted by the Dirac mass term. We conclude in \S\ref{sec:conclusions}. 
Two Appendices provide more details on the numerical techniques used to solve for the zero modes using Chebyshev interpolation, and on the derivation of the low-energy effective action.

\section{Axion Strings} \label{sec:massless}

    In this section, we will review the original ultraviolet completion of an axion string studied by Callan and Harvey~\cite{Callan:1984sa}, with Lagrangian 
    \begin{equation}
        \mathcal{L} = -\frac{1}{4} F_{\mu \nu} F^{\mu \nu} + \bar{\psi} i \slashed{D} \psi + |\partial_\mu \Phi|^2 + y \es \bar{\psi} \!\es \left(\Phi_1 + i \gamma^5 \Phi_2\right)\!\es\psi - V(\Phi)\,. \label{eq:noMassLag}
    \end{equation}
    This theory lives in four-dimensional Minkowski space $\mathcal{M}_4$ and consists of an abelian gauge field $A_\mu$ with field strength $F_{\mu \nu} = \partial_\mu A_\nu - \partial_\nu A_\mu$, an uncharged complex scalar field $\Phi \equiv \Phi_1 + i \Phi_2 \equiv f(x) \e^{i \theta(x)}$ with potential
    \begin{equation}
        V(\Phi) = \lambda\left(|\Phi|^2 - v^2\right)^2\,, \label{eq:scalarFieldPotential}
    \end{equation}
    and a single charged Dirac fermion $\psi$, which chirally couples to $\Phi$ with strength $y$. We use $D_\mu = \partial_\mu - i e A_\mu$ to denote the gauge covariant derivative, where $\slashed{D} \equiv \gamma^\mu D_\mu$, and $\gamma^\mu$ are the standard Dirac gamma matrices with $\{\gamma^\mu, \gamma^\nu\} = 2 \eta^{\mu \nu}$ and $\gamma^5 \equiv i \gamma^0 \gamma^1 \gamma^2 \gamma^3$. This theory enjoys a $\lab{U}(1)_\slab{pq}$ global symmetry, commonly called the Peccei-Quinn symmetry, under which the complex scalar and fermion transforms as $\Phi \to \e^{i \alpha} \Phi$ and $\psi \to \e^{\sminus i \gamma^5 \alpha/2} \psi$, respectively.

    \subsection{The Axion String}
        The potential (\ref{eq:scalarFieldPotential}) forces the scalar $\Phi$ to acquire a vacuum expectation value $\langle \Phi \rangle = v$. It also allows for the existence of static axionic string configurations,
        \begin{equation}
            \Phi_n(x) = f(r) \es \e^{i n \varphi}\,, \label{eq:axStringSol}
        \end{equation}
        characterized by an integer topological charge $n \in \mathbb{Z}$. We work in standard cylindrical coordinates $(x, y, z) = (r \cos \varphi, r \sin \varphi, z)$, with the string oriented along the $z$-axis. In these configurations, the phase of the complex scalar field, i.e. the ``axion'' $\theta(x) \equiv \lab{arg} \, \Phi(x)$, winds $n$ times around its field space $\theta \sim \theta + 2 \pi$ as we move around the string. Said differently,
        \begin{equation}
            n \equiv \frac{1}{2 \pi} \oint_{\gamma} \ud \theta\,,
        \end{equation}
        where $\gamma$ is a closed contour that encircles the string at $r = 0$. The ``radial profile'' of this string is determined by the real function $f(r)$. Far from the string, the potential (\ref{eq:scalarFieldPotential}) forces the scalar to sit in the minimum of its potential and so $f(r) \to v$ as $r \to \infty$. Likewise, regularity of the solution forces $f(r)$ to vanish in the core of the string: $f(r) \to 0$ as $r \to 0$. These solutions are also called ``global vortices'' in the literature, since they are vortex solutions that are charged under the \emph{global} $\lab{U}(1)_\slab{pq}$ symmetry, as opposed to a gauge symmetry.

        The tension of an axion string with charge $n$ is given by
        \begin{equation}
            T_n = 2\pi \int_0^{L}\!\ud r\, r \left[ \left(\frac{\ud f}{\ud r}\right)^2 \! + \frac{n^2}{r^2}f^2 + \lambda(f^2 - v^2)^2\right]\,. \label{eq:bgStringTension}
        \end{equation}
        Since $f(r) \to v$ as $r \to \infty$, the tension diverges logarithmically with the size of the system or IR cutoff $L$,
        \begin{equation}
            T_n \sim 2 \pi v^2 n^2 \log \left(\frac{L}{r_\lab{core}}\right)\,,\mathrlap{\qquad L \to \infty\,.}
        \end{equation}
        Here, we have introduced the ``size'' of the string $r_\lab{core}$, which is of order $r_\lab{core} = \big(\sqrt{\lambda} v\big)^{\sminus 1}$. This IR divergent total energy is common for global strings and is typically regulated by either placing the theory in a box or by positing that there is another string of opposite charge a distance $L$ away, as is often the case in, e.g. cosmological simulations of axion string networks. 

\begin{figure}
            \centering
            \includegraphics{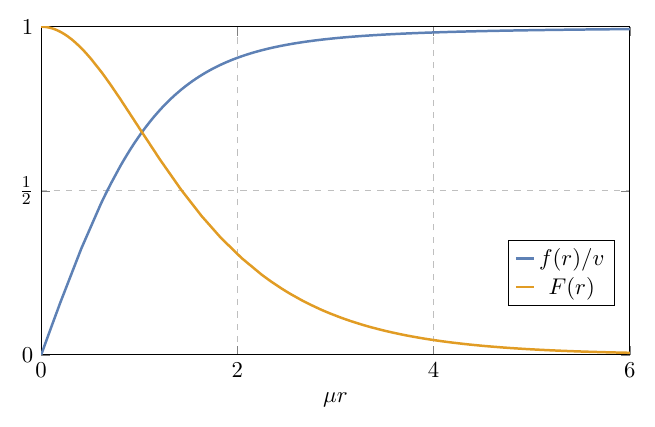}
            \caption{The radial profile $f(r)$ [{\color{Mathematica1} blue}], in units of $v$, for the axion string of charge $\pm 1$ and corresponding (non-normalized) zero mode profile $F(r)$ [{\color{Mathematica2} orange}] with $y = \sqrt{\lambda}$. \label{fig:masslessComb}}
        \end{figure}

        Minimizing the tension yields the equation of motion 
        \begin{equation}
            \frac{1}{r} \frac{\ud}{\ud r}\!\left(r \es\frac{\ud f}{\ud r}\right) - \left[\frac{n^2}{r^2} + 2 \lambda (f^2 - v^2 )\right] f = 0\,. \label{eq:bgStringEom}
        \end{equation}
        In general, this equation cannot be solved analytically but is amenable to numerics. We will restrict our attention throughout to strings of charge $n = +1$. An asymptotic analysis of (\ref{eq:bgStringEom}) shows that the radial profile $f(r)$ behaves as
        \begin{equation}
            f(r) \sim\begin{dcases}
                v \es \es C_1\es \big(\sqrt{\lambda} v r\big) + \cdots & r \to 0 \\
                v \! \left(1 - \frac{1}{4 \lambda v^2 r^2}\right) + \cdots & r \to \infty
            \end{dcases}\,, \label{eq:radialProfileAsymptotics}
        \end{equation}
        with $C_1$ an overall constant that can be analytically determined by matching the two solutions in an intermediate region, or by numerics. It is clear from the behavior as $r \to \infty$ that the string solution varies over length scales $r_\lab{core}= \big(\sqrt{\lambda} v\big)^{\sminus 1}$. We may then use the techniques outlined in Appendix~\ref{app:numerics} to numerically search for the solution which varies over scales set by $r_\lab{core}$ with the correct asymptotics (\ref{eq:radialProfileAsymptotics}). We show this profile in Figure~\ref{fig:masslessComb}.

    \subsection{Fermionic Zero Modes} \label{sec:fzm}
        
        We are interested in the low energy dynamics of (\ref{eq:noMassLag}) in the background of the axion string. Absent the string, the fermion acquires a Dirac mass $\mu = y v$---which we call the ``Yukawa mass'' since it arises from the spontaneous breaking of the $\lab{U}(1)_\slab{pq}$ symmetry---and so the low-energy theory only contains the massless axion and the abelian gauge field. One might think that the low-energy theory in the presence of the axionic string is similar; far from the string, the fermions again get a mass $\mu$ and the low-energy theory should be unchanged. However, as argued in \cite{Callan:1984sa} this reasoning is incomplete: there are instead massless chiral zero modes which are localized to the string. In this section, we will review the direct construction of these zero modes. A common intuitive explanation is that these massless modes exist because the radially-dependent ``mass'' of the bulk fermions $y f(\rho)$ vanishes in the core of the string. However, we argue in Section~\ref{sec:massive} that this is not a necessary condition, while in the Section~\ref{sec:inflow} we discuss why these modes are necessary from the perspective of anomaly inflow.

        In the presence of the string of charge $n = +1$ and no background gauge field, the fermions obey the equation of motion
        \begin{equation}
            \left( i \slashed{\partial} + y \es  f(r) \es \e^{i \gamma^5 \varphi}\right) \! \psi = 0\,. \label{eq:zmEom}
        \end{equation}
        For simplicity, we will restrict to solutions of the form
        \begin{equation}
            \psi(x) = \psi(r, \varphi)\, \e^{\sminus i p(t + z)}\,,
        \label{eqn:zeromode}
        \end{equation}
        where $\psi(r, \varphi) = \psi_\alpha(r, \varphi)$ is a four-component spinor function with $\alpha = 0, \ldots, 3$. This ansatz describes solutions that travel in the $(\minus z)$-direction at the speed of light. In the Weyl representation, (\ref{eq:zmEom}) reduces to the set of coupled equations
        \begin{equation}
            \begin{aligned}
                0 &= 2p \es \psi_2 + \e^{\sminus i \varphi} \left(y  f(r)\es \psi_0 + i \partial_r \psi_3 + r^{\sminus 1} \partial_\varphi \psi_{3}\right) \\
                0 &=  y f(r) \e^{\sminus i \varphi}\es \psi_1 + \e^{i \varphi} \left(i \partial_r \psi_2 - r^{\sminus 1} \partial_\varphi \psi_2\right) \\
                0 &=  y f(r) \e^{i \varphi} \es  \psi_2 - \e^{\sminus i \varphi} \left(i \partial_r \psi_1 + r^{\sminus 1} \partial_\varphi \psi_1\right) \\
                0 &=  2p \es \psi_1 +\e^{i \varphi} \left(y f(r) \es \psi_3 - i \partial_r \psi_0 + r^{\sminus 1} \partial_\varphi \psi_0\right)
            \end{aligned}\,.
        \end{equation}
        Such solutions should have definite helicity, and so we may set $\psi_1 = \psi_2 = 0$. We then see that $\psi_0 = \psi_0(r)$ and $\psi_3 = \psi_3(r)$ are purely radial functions that satisfy the coupled system of equations
        \begin{equation}
            \begin{aligned}
                0 &= y f(r)\es \psi_0 + i \partial_r \psi_3 \\ 
                0 &= y f(r) \es \psi_3 - i \partial_r \psi_0 
            \end{aligned}\,. \label{eq:masslessZMEom}
        \end{equation}
        These equations can be easily solved to find two solutions of the form
        \begin{equation}
            \psi_p(x) = \sqrt{\frac{p}{2}} \begin{psmallmatrix} 1 \\ 0 \\ 0 \\  \mp i \end{psmallmatrix} \e^{\sminus i p(t + z)}F(r)\,,
            \label{eq:masslessZM}
        \end{equation}
        where $F'(r) = \pm y f(r) F(r)$ or
        \begin{equation}
            F(r) =  \mathcal{A} \exp\left(\pm y \int_{0}^{r}\!\ud r' \, f(r')\right), \label{eq:fDef}
        \end{equation} 
        with $\mathcal{A}$ an overall constant of integration. For this solution to be appropriately normalizable, 
        \begin{equation}
            \int_{\mathbb{R}^3} \!\ud^3 x\, \psi^\dagger_{p}(x) \psi_{p'}(x) = 2\pi |p|\es\es  \delta(p - p')\,, \label{eq:normalizationCond}
        \end{equation}
        or specifically
        \begin{equation}
            2\pi \int_{0}^{\infty} \! \ud r\, r\es\es F^2(r) = 1\,,
        \end{equation}
        so we must select the negative sign in (\ref{eq:fDef}). We thus find that the axion string with charge $+1$ supports a chiral fermionic zero mode that travels at the speed of light in the $(\minus z)$-direction, with form
        \begin{equation}
            \psi(x) = \mathcal{A}\sqrt{\frac{p}{2}} \begin{psmallmatrix} 1 \\ 0 \\ 0 \\  \sminus i \end{psmallmatrix} \e^{\sminus i p(t + z)}  \exp\left(\minus y \int_0^{r}\!\ud r'\, f(r')\right)\,.
        \end{equation}
        If we were to repeat this analysis with the charge $\minus 1$ axion string, then it would support a normalizable chiral zero mode that instead travels at the speed of light in the $(+z)$-direction,
        \begin{equation}
            \psi(x) = \mathcal{A}\sqrt{\frac{p}{2}} \begin{psmallmatrix} 0 \\ 1 \\ \sminus i \\  0 \end{psmallmatrix} \e^{\sminus i p(t - z)}  \exp\left(\minus y \int_0^{r}\!\ud r'\, f(r')\right)\,.
        \end{equation}
        In Figure~\ref{fig:masslessComb}, we show the radial profile $F(r)$ for $y = \sqrt{\lambda}$. These zero modes are localized in a region about the string of size $r_\slab{zm} = r_\lab{core}/y$.

        Before we move on, it will be useful to understand why the presence of the axion string permits these normalizable modes to exist, solely from the perspective of the differential equations. We will see that, contrary to naive expectation, it is not because the fermion is ``massless'' in the core of the string. With (\ref{eqn:zeromode}), we can decouple the equations in (\ref{eq:masslessZMEom}) to find two copies of the second order equation
        \begin{equation}
            \left[\frac{\ud^2}{\ud r^2} - \frac{f'(r)}{f(r)} \frac{\ud}{\ud r} - y^2 f^2(r)\right]\! F(r) = 0\,, \label{eq:zm2Eom}
        \end{equation}
        with $\psi_0(r) = F(r)$ and $\psi_3(r) = \minus i F(r)$. As is clear from Figure~\ref{fig:masslessComb}, the function $f(r)$ is smooth and non-singular for all $r \in (0, \infty)$. This implies that the solutions to (\ref{eq:zm2Eom}) are regular everywhere, except possibly as $r \to 0$ and $r \to \infty$, which are regular and irregular singular points, respectively. A normalizable solution must be regular at both of these points and generally fails to exist because the solution that is regular about one singularity is not regular about the other.

        Using the asymptotics (\ref{eq:radialProfileAsymptotics}) of the background $f(r)$, far from the string (\ref{eq:zm2Eom}) reduces to
        \begin{equation}
            \left[\frac{\ud^2}{\ud r^2} - \mu^2 + \cdots \right] \! F(r) = 0\,,
        \end{equation}
        which has solutions that behave as
        \begin{equation}
            F(r) = C_\subm F_\subm(r) + C_\subp F_\subp(r) \sim C_\subm \es \e^{\sminus \mu r} + C_\subp \es \e^{\mu r}\,,\mathrlap{\qquad r \to \infty\,,}
        \end{equation}
        where we have defined two linearly-independent solutions $F_{\pm}(r)$ with definite scaling as $r \to \infty$.
        In this limit, the fermion zero modes do not ``see'' the string, but merely feel their acquired mass $\mu = y v$. However, near the core of the string, (\ref{eq:zm2Eom}) instead reduces to
        \begin{equation}
            \left[\frac{\ud^2}{\ud r^2} - \frac{1}{r} \frac{\ud}{\ud r} + \cdots\right]\!F(r) = 0\,. 
        \end{equation}
        As we move towards the core of string $r \to 0$, there is a balance between the first and second terms in (\ref{eq:zm2Eom}), the latter of which only appears due to the existence of the string. The third term, due to the Yukawa interaction, can be neglected. Solutions then behave as
        \begin{equation}
            F(r) = C_1 F_1(r) + C_2 F_2(r)  \sim C_1 + C_2 r^2\,,\mathrlap{\qquad r \to 0\,,}
        \end{equation}
        where we have again identified two solutions $F_{1,2}(r)$ which have definite scaling as $r \to 0$. Importantly, both of these solutions are regular near the origin.

        We can understand why the axion string allows a normalizable zero mode to exist as follows. Since (\ref{eq:zm2Eom}) is a second-order differential equation, it has only two linearly-independent solutions. Above, we found that a general solution can either be expressed as a linear combination of $F_{1, 2}(r)$ or $F_\pm(r)$. Since these two sets of solutions are not linearly independent, there always exists a linear map between them. For a solution to be normalizable, it must decay as $r \to \infty$ and so we must have that $C_\subp = 0$; therefore we must be able to write this solution as $F(r) = C_\subm F_\subm = C_{\subm, 1} F_1(r) + C_{\subm, 2} F_2(r)$. It is often the case that only one of the solutions $F_1(r)$ or $F_2(r)$ is well-behaved at the origin $r = 0$ and, unless there is some special structure that ensures it, the solution that is well-behaved as $r \to \infty$ will not be well-behaved as $r \to 0$. Fortunately, the presence of the string ensures that \emph{both} linearly-independent solutions are regular as $r \to 0$. Since $f(r)$ is a smooth function, we are also guaranteed that the solutions are regular for positive $r$. Consequently, a normalizable solution to (\ref{eq:zm2Eom}) exists, regardless of its exact form.

        An analogous situation occurs for the unbound states of the quantum mechanical hydrogen atom, whose radial modes obey
        \begin{equation}
            \left[-\frac{1}{2 r^2} \frac{\ud}{\ud r} \left(r^2 \frac{\ud}{\ud r}\right) + \frac{\ell(\ell+1)}{2r^2} - \frac{\alpha}{r} - E \right] \!\psi(r) = 0\,.
        \end{equation}
        Due to the centrifugal barrier, solutions behave as $\psi(r) \sim C_1 r^\ell + C_2 r^{-(\ell+1)}$ as $r \to 0$ and $\psi(r) \sim C_\subm \e^{\sminus \sqrt{\sminus 2 E} r} + C_\subp \e^{\sqrt{\sminus 2 E}r}$ as $r \to \infty$. Bound states, $E < 0$, have a discrete spectrum because it is impossible to simultaneously impose that the wavefunction be both regular at the origin \emph{and} exponentially decay at spatial infinity, except at a discrete set of energy eigenvalues. For all other energies, solutions with $C_\subp = 0$ necessarily have $C_2 \neq 0$. For unbound states, $E > 0$, there is no restriction on the behavior as $r \to \infty$, and so there exists a continuum of regular solutions with $C_2 = 0$, for all positive energies.

        Having reviewed how axion strings, and their associated fermionic zero modes, arise in the Callan--Harvey model (\ref{eq:noMassLag}), we are now in a position to understand how these zero modes behave as we deform (\ref{eq:noMassLag}). In the next section, we will study what happens to these zero modes in the presence of a Dirac mass $m$ for the four-dimensional fermion. Even though the four-dimensional fermion is everywhere massive, these fermionic zero modes still exist as long as $|m| < \mu$. Furthermore, we argue that there is an interesting ``phase structure'' in which the zero modes become increasingly unbound from the string as $|m| \to \mu$, eventually disappearing for $|m| > \mu$.

\section{Adding a Mass} \label{sec:massive}
	
	In the previous section, we reviewed how fermionic zero modes could arise along an axion string in the context of a simple toy model. In that specific UV completion, the axion string is a solitonic object in which the $\lab{U}(1)_\slab{pq}$ symmetry that is spontaneously broken far from the string is restored at the core. This causes the four-dimensional fermion to see effectively zero mass near the core, and a popular refrain is that this why we should expect to see massless fermionic zero modes localized on the string. The goal of this section is to explain why this is not a necessary condition and see that fermionic zero modes can arise even in theories where bulk four-dimensional fermions have a nowhere vanishing mass.

	We will deform (\ref{eq:noMassLag}) by adding a mass $m$ to the Dirac fermion $\psi$,
	\begin{equation}
		\mathcal{L} = -\frac{1}{4} F_{\mu \nu} F^{\mu \nu} + \bar{\psi} (i \slashed{D} - m) \psi + |\partial_\mu \Phi|^2 + y \es \bar{\psi} \left(\Phi_1 + i \gamma^5 \Phi_2\right)\psi - V(\Phi)\,, \label{eq:massLag}
	\end{equation}
	again with $V(\Phi) = \lambda(|\Phi|^2 - v^2)$. We restrict to the case where the amplitude of $\psi$ is small. This theory permits the same axion string solution (\ref{eq:axStringSol}) as before. However, the fermion now obeys a modified equation,
	\begin{equation}
		\left(i \slashed{\partial} - M(r, \varphi)\es \e^{i \gamma^5 \alpha(r, \varphi)} \right)\!\psi = 0\,, \label{eq:zmEomMass}
	\end{equation}
	where $M(r, \varphi) \es \e^{i \gamma^5 \alpha(r, \varphi)} = m - y \es f(r) \es \es \e^{i \gamma^5 \varphi}$ or, explicitly, 
	\begin{equation}
		\begin{aligned}
			M(r, \varphi) &= \sqrt{\big(m - y f(r) \cos \varphi\big)^2 + y^2 f^2(r) \sin^2 \varphi } \\
			\alpha(r, \varphi) &= \lab{arg}\es\es \big(m - y f(r)\es \e^{\sminus i \varphi}\big) = i \log \! \left(\frac{m - y f(r) \e^{\sminus i \varphi}}{M(r, \varphi)}\right)
		\end{aligned}\,\,. \label{eq:massMass}
	\end{equation}
    In this case, the fermion mass $M(r, \varphi)$ no longer goes to zero at the core of the string even though $f(r) \to 0$, but instead approaches the ``core mass'' $ M(r = 0, \varphi) = m$. Even so, we will still find that this string supports fermionic zero modes as long as this core mass is less than the Yukawa mass, $|m| < \mu$.

	To find these zero modes, we again restrict to axion strings with charge $+1$ and search for zero modes of the form (\ref{eqn:zeromode}). The mass deformation $m \bar{\psi} \psi$ explicitly breaks the chiral symmetry or, analogously, the continuous axion shift symmetry $\varphi \to \varphi + \alpha$, and thus changes the form of the axion string solution. However, classically this potential is controlled by fermion's number density $\bar{\psi} \gamma^0 \psi$, and in the limit of small amplitude---or, quantum mechanically, small occupation numbers---this effect is subleading and can be ignored. Similarly, we will not consider the electromagnetic field generated by the charged fermion, setting the vector potential $A_\mu = 0$, as this effect is also subleading in the limit of small amplitudes. 

    It is worth noting that integrating out the fermion generates a potential for the scalar that depends on $\varphi$. The leading Coleman--Weinberg estimate $\propto M^4(r,\varphi) \log(M(r,\varphi)/ \mu)$~\cite{Coleman:1973jx, Agrawal:2023sbp} is minimized along $\varphi = 0$ and leads to a domain wall for the axion that emanates off the string and extends in this direction. In what follows, we are explicitly assuming that this contribution is fine-tuned away, though we will see that there is a remnant of the axion domain wall in the orientation of the zero mode profiles.

    \begin{figure}
        \centering
        \includegraphics[trim={0cm 1.4cm 0 2.3cm}, clip, width=\textwidth]{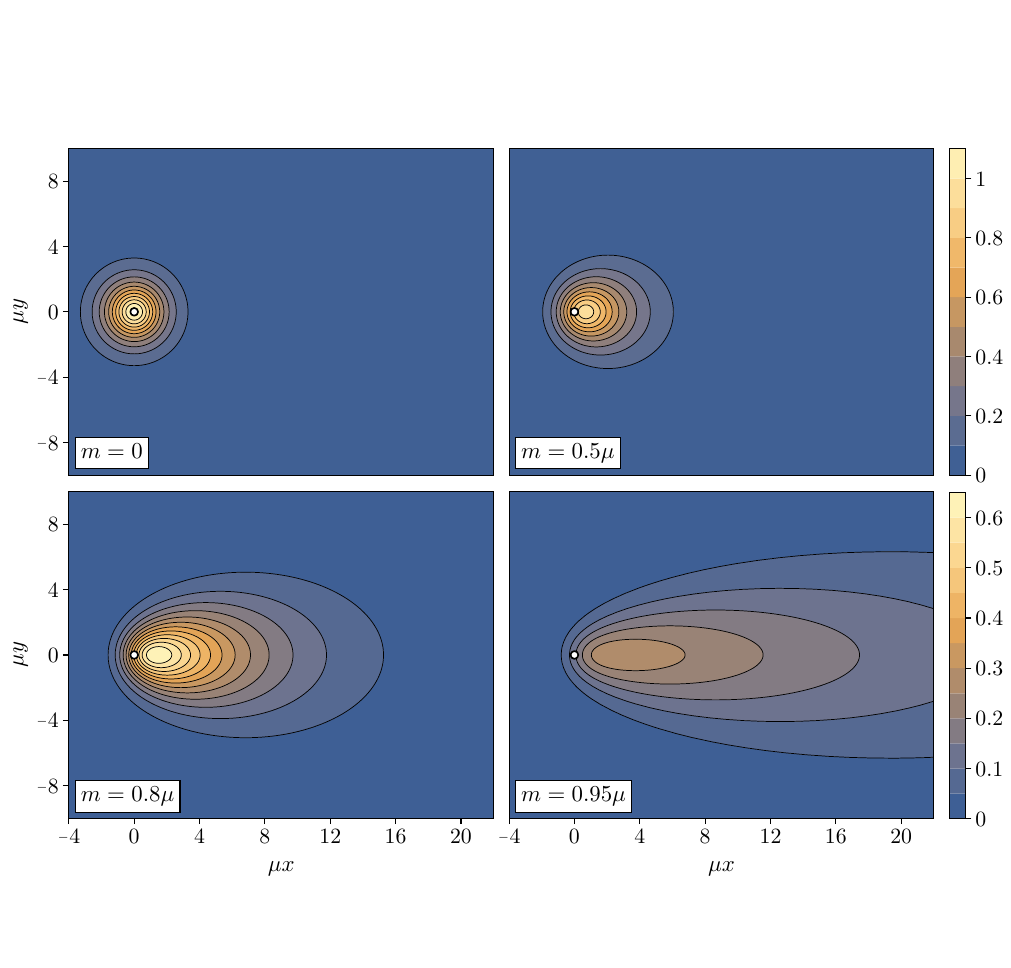}
        \caption{The profile functions of the zero mode wavefunction $|\psi_0(r, \varphi)|$ in the presence of a Dirac core mass $m$ and normalized according to (\ref{eq:normalizationCond}), for various values of $m/\mu$ and $\sqrt{\lambda} = y = \sqrt{0.1}$, with contours in units of $\sqrt{\omega}\mu$. The other component satisfies $\psi_3(r, \varphi) = \minus i \psi_0(r, \minus \varphi)$ and is thus identical in structure. The white dot denotes the origin of the string, $r = 0$. As $m \to \mu$, the zero mode profiles spread out and become unbound from the string---note the difference in scales of the contours between the two rows. For $m > \mu$, these zero modes cease to exist.  Qualitatively similar results apply when $\sqrt{\lambda}\neq y$. \label{fig:massiveZeroModePlot}} 
    \end{figure}

    By Lorentz symmetry, we can set $\psi_1(r, \varphi) = \psi_2(r, \varphi) = 0$ since the modes we are searching for propagate in the $(\minus z)$-direction at the speed of light, and so in the Weyl representation (\ref{eq:zmEomMass}) reduces to
	\begin{equation}
		\begin{aligned}
			0 &= -m \psi_0 + \e^{\sminus i \varphi}\! \left(y  f(r)\es \psi_0 + i \partial_r \psi_3 + r^{\sminus 1} \partial_\varphi \psi_{3}\right) \\
			0 &=  - m \psi_3 +\e^{i \varphi} \! \left(y f(r) \es \psi_3 - i \partial_r \psi_0 + r^{\sminus 1} \partial_\varphi \psi_0\right) \label{eq:zmEomMassComp}
		\end{aligned}\,.
	\end{equation}
    We can solve (\ref{eq:zmEomMassComp}) numerically using Chebyshev interpolation, whose details we describe in Appendix~\ref{app:numerics}. The resulting profile functions are shown in Figure~\ref{fig:massiveZeroModePlot} for various values of $m/\mu$. In the plot, we see the normalized zero mode as a function of dimensionless $x$ and $y$ variables, with a white dot corresponding to the location of the origin and hence the axion string. Surprisingly, we find that this zero mode becomes less localized to the string the larger the core mass $m$ is. For large $m \lesssim \mu$, it is no longer centered at the core of the axion string and instead stretches out along the positive $x$-axis. This direction is determined by the orientation of the string and specifically the axis along which $\varphi = 0$, corresponding to the minimum of $M(r, \varphi)$. As $m \to \mu$, we find that there is a sort of phase transition in which case the zero mode completely delocalizes onto a two-dimensional wedge. Finally, the mode completely disappears for $m > \mu$.

	To better understand the counterintuitive results shown in Figure~\ref{fig:massiveZeroModePlot}, in \S\ref{sec:pert} we first solve (\ref{eq:zmEomMassComp}) by treating $m$ as a small perturbation. As in \S\ref{sec:fzm}, we will see that it is the presence of the axion string, and not a vanishing mass at the core, which allows for this zero mode to exist. In \S\ref{sec:domainWalls}, we describe a simple analog of this mechanism for Dirac fermions in the three-dimensional half-space~$\mathbb{R}^{2, 1}_+$. In section \S\ref{sec:critMass}, we analyze the zero mode for the critical case $m = \mu$.

	\subsection{The Small Mass Limit} \label{sec:pert}

		In this section, we will argue that a normalizable zero mode solution to (\ref{eq:zmEomMassComp}) exists in the small Dirac mass limit by analyzing the structure of (\ref{eq:zmEomMassComp}) when $m \ll \mu$. Expanding each spinor into Fourier modes,
		\begin{equation}
			\psi_{\alpha}(r, \varphi) = \sum_{\ell \in \mathbb{Z}} \psi_{\alpha, \ell}(r) \e^{i \ell \varphi}\,,
			\label{eq:FourierExpansion}
		\end{equation}
		the zero mode equation (\ref{eq:zmEomMassComp}) reduces to an infinite set of coupled equations
		\begin{equation}
			\begin{aligned}
				0 &= -m \psi_{0, \ell-1} + y  f(r)\es \psi_{0,\ell} + i \partial_r \psi_{3,\ell} + i \ell r^{\sminus 1}\psi_{3,\ell} \\
				0 &=  - m \psi_{3, \ell+1} + y f(r) \es \psi_{3,\ell} - i \partial_r \psi_{0,\ell} + i \ell r^{\sminus 1} \psi_{0,\ell}\, .
			\end{aligned}\, \label{eq:zmEomMassFourier}
		\end{equation}
		When $m = 0$, the length scale over which the zero mode varies is set by the Yukawa mass $\mu = y v$, which dominates the system of equations (\ref{eq:zmEomMassFourier}) as $r \to \infty$. The mass $m$ will also be relevant as $r \to \infty$, and so our perturbative expansion will be in powers of $m/\mu$.

		We will thus search for solutions of the form 
		\begin{equation}
		\label{eq:massive_zero_mode_ansatz}
			\psi(x) = \sqrt{\frac{p}{2}}  \left[\begin{psmallmatrix} 1 \\ 0 \\ 0 \\  \sminus i \end{psmallmatrix} F(r) + \delta \psi(r, \varphi)\right]\e^{\sminus i p(t + z)}\,,
		\end{equation}
		and work to first order in $\delta \psi = \mathcal{O}(m/\mu)$. In this limit, (\ref{eq:zmEomMassFourier}) reduces to two identical sets of differential equations,
		\begin{equation}
			\begin{aligned}
				\left(\frac{\ud}{\ud r} - \frac{1}{r}\right) \!G(r) &= y f(r) H(r) \\
				\left(\frac{\ud}{\ud r} + \frac{1}{r}\right) \! H(r) &= y f(r) G(r) - m F(r)
			\end{aligned}\,,
		\end{equation}
		where $(G, H) = (\delta \psi_{0, 1}, \minus i \delta \psi_{3,1})$ or $(G, H) = ( i \delta \psi_{3, \sminus 1}, \minus \delta \psi_{0, \sminus 1})$. All other equations in (\ref{eq:zmEomMassFourier}) are trivially satisfied by setting the other $\delta \psi_{\alpha, \ell}(r) = 0$. The function $G(r)$ satisfies the inhomogeneous equation
		\begin{equation}
			\begin{aligned}
				\left[\frac{\ud^2}{\ud r^2} - \frac{f'(r)}{f(r)} \frac{\ud}{\ud r} + \frac{f'(r)}{r f(r)} - y^2 f^2(r)\right] G(r) &= - m y f(r) F(r)\,, \label{eq:gEq}
			\end{aligned}
		\end{equation}
		which can be solved using solutions to the homogeneous equation (with $m = 0$) via variation of parameters~\cite{Bender:1999amm}. These homogeneous solutions will determine how the inhomogeneous solution behaves, and so it will be helpful to understand their asymptotic behavior.

		Far from the string, at $r \to \infty$, the homogeneous equation reduces to
		\begin{equation}
			\left[\frac{\ud^2}{\ud r^2} - \mu^2 + \cdots \right]\!G(r) =  0\,,
		\end{equation}
		in which case the general solution behaves as 
		\begin{equation}
			G(r) = C_\subm G_\subm(r) + C_\subp G_\subp(r) \sim C_\subm \e^{\sminus \mu r} + C_\subp \e^{\es \mu r}\,, \mathrlap{\qquad r \to \infty\,,}
		\end{equation}
		where we have again introduced the two linearly-independent solutions $G_\pm(r)$ with definite scaling as $r \to \infty$. We will choose their normalization such that the Wronskian is
		\begin{equation}
			W(r) = 2 y f(r)\,.
		\end{equation}
		Likewise, as $r \to 0$, the homogeneous equation simplifies to
		\begin{equation}
			\left[\frac{\ud^2}{\ud r^2} - \frac{1}{r} \frac{\ud}{\ud r} + \frac{1}{r^2} + \cdots \right] \!G(r) = 0\,.
		\end{equation}
		As before, the presence of the axion string introduces a regular singularity at $r = 0$, and solutions take the form
		\begin{equation}
			G_\spm(r) =  C_{\spm, 1} \sum_{k = 0}^{\infty} a^{\spm}_k r^{k+1} + C_{\spm,2} \sum_{k = 0}^{\infty} b^{\spm}_k r^{k +1} \log r\,. \label{eq:gFrob}
		\end{equation}
		Crucially, both of the linearly-independent homogeneous solutions are regular as $r \to 0$, regardless of the overall coefficients $C_{\spm, 1}$ or $C_{\spm, 2}$. The most general solution to (\ref{eq:gEq}) that decays as $r \to \infty$ is
		\begin{equation}
			G(r) = \frac{1}{2} m\es C_\subm G_\subm(r) + \frac{1}{2} m \, G_\subp(r) \!\int^{\infty}_r \!\ud r'\, F(r') \es G_\subm(r') + \frac{1}{2} m \, G_\subm(r) \! \int^{r}_0 \! \ud r'\, F(r') \es G_\subp(r')\,, \label{eq:gSol}
		\end{equation}
		for arbitrary values of $C_\subm$. Since $F(r)$ and $G_\subm(r)$ both decay as $\e^{\sminus \mu r}$, while $G_\subp(r')$ grows as $\e^{\es \mu r}$, all three terms in (\ref{eq:gSol}) decay as $\e^{\sminus \mu r}$ as $r \to \infty$. Likewise,  both $F(r)$ and the homogeneous solutions $G_{\pm}(r)$ are regular as $r \to 0$, and thus so is (\ref{eq:gSol}). This solution is thus normalizable.

		Given a solution for $G(r)$, we can solve for the other profile function $H(r)$ via
		\begin{equation}
			H(r) = \frac{1}{y f(r)}\left(\frac{\ud}{\ud r} - \frac{1}{r}\right) \!G(r)\,. \label{eq:heq}
		\end{equation}
		This equation should cause some alarm: since $f(r) \propto \sqrt{\lambda}v r$ diverges as $r \to 0$, we might worry that $H(r)$ is \emph{not} regular as $r \to 0$ even though $G(r)$ is. However, the free coefficient $C_\subm$ in (\ref{eq:gSol}) can generally be chosen to yield a normalizable $H(r)$. Writing (\ref{eq:heq}) as
		\begin{equation}
			\begin{aligned}
				H(r) &= \frac{m}{2y f(r)} \Bigg[ C_\subm \, r \es \frac{\ud}{\ud r}\!\left(\frac{G_\subm(r)}{r}\right) +  r \es \frac{\ud}{\ud r}\!\left(\frac{G_\subp(r)}{r}\right)  \int_{0}^{\infty}\!\ud r'\, F(r') G_\subm(r')   \\
				&\qquad - r \es \frac{\ud}{\ud r}\!\left(\frac{G_\subp(r)}{r}\right)  \int_{0}^{r}\!\ud r'\, F(r') G_\subm(r') +   r \es \frac{\ud}{\ud r}\!\left(\frac{G_\subm(r)}{r}\right) \! \int_{0}^{r}\!\ud r'\, F(r') G_\subp(r')\Bigg]
			\end{aligned}\,\,, \label{eq:hSol}
		\end{equation}
		the combinations
		\begin{equation}
			r\frac{\ud}{\ud r}\!\left(\frac{G_\pm(r)}{r}\right) =  C_{\spm, 1} \sum_{k =0}^{\infty} a^\pm_k k r^{k} +  C_{\spm, 2}\sum_{k = 0}^{\infty} b^\pm_k (1 + k \log r) r^k\,,
		\end{equation}
		are crucially regular as $r \to 0$, approaching a constant $b^\pm_0$. 
		Furthermore, since each of the integrals in the second line of (\ref{eq:hSol}) decay as $\propto r^2 \log r$ as $r \to 0$, each of those terms vanish as $r \to 0$ and thus do not generate a non-normalizable contribution to $H(r)$. 

		In contrast, the terms in the first line are dangerous: they can contain a constant piece which, when divided by $f(r)$, would cause $H(r) \sim H_0/r$ as $r \to 0$ for some constant $H_0$. Note, however, that we can always\footnote{The only way this strategy could fail is if both $G_\pm(r)$ also have definite scaling behavior as $r \to 0$, such that either $a_k^\subp = b_k^\subm = 0$. In this case, we cannot tune $C_\subm$ to cancel the constant and the solution is non-normalizable. While this behavior is extremely non-generic, we would need the connection formulae for (\ref{eq:gEq}) to prove it does not happen. We take our numerical solutions to be proof that it does not.} choose $C_\subm$ so that the overall constant term in the square braces of (\ref{eq:hSol}) cancels. In this case, $H(r) \sim H_0 \log r$ as $r \to 0$, for some constant $H_0$. But, this is still square-integrable, and so we conclude that we can construct, at least perturbatively, a normalizable zero mode solution in the presence of a small, non-zero Dirac mass $m$. As in the unperturbed case reviewed in Section~\ref{sec:massless}, the axion string modifies the wave equation so that both linearly-independent solutions are regular as $r \to 0$, allowing for a normalizable zero mode.

        The numerical solutions presented in Figure~\ref{fig:massiveZeroModePlot} demonstrate that these solutions exist non-perturbatively as well, as long as $m < \mu$. As $m \to \mu$, these solutions become unbound from the string and occupy a two-dimensional ``wedge'' in the plane orthogonal to the string. To better understand the critical case $m = \mu$, and why these solutions exist at all, it will be helpful to first study a simpler, albeit analogous, system in which normalizable zero modes exist even in the presence of a nowhere vanishing gap.

	\subsection{Zero Modes on Domain Walls} \label{sec:domainWalls}

		That a massless field can emerge from one that is everywhere massive is counterintuitive, so it will be helpful to consider a simpler system where it also occurs: the Dirac fermion on the half-plane $\mathbb{R}_\subp^{2,1}$~\cite{Callan:1984sa,Kaplan:1992sg,Aitken:2017nfd,Fradkin:2013sab}. Studying this system\footnote{This system naturally appears, for instance, in the study of the quantum anomalous hall effect~\cite{Haldane:1988zza} along interfaces between topological and regular insulators~\cite{Fradkin:2013sab}.} will also help us understand why these zero modes completely delocalize onto a two-dimensional wedge when $m = \mu$, which we will discuss in detail in \S\ref{sec:critMass}.

		Let us first consider a single Dirac fermion in the full space $\mathbb{R}^{2,1}$ with a spatially-varying mass, and Lagrangian
		\begin{equation}
			\mathcal{L} = \bar{\psi}\!\left(i \slashed{\partial} - m(x)\right)\!\psi\,. \label{eq:3dDirac}
		\end{equation}
		Here, $(\gamma^0, \gamma^1, \gamma^2) = (\sigma_1, i \sigma_3, \minus i \sigma_2)$ are the three-dimensional $\gamma$-matrices and we use the Cartesian coordinates $(t, x, z)$. We take the mass term $m(x)$ to only depend on the coordinate $x$. We can search for zero modes that move at the speed of light in the $(+z)$-direction by assuming that there exists a solution of the form
		\begin{equation}
			\psi(x) = \sqrt{p} \, \e^{\sminus i p(t - z)} \! \begin{pmatrix} \eta(x) \\ \chi(x) \end{pmatrix}\,.
		\end{equation}
        with $\eta(x)$ and $\chi(x)$ functions of $x$. With this ansatz, (\ref{eq:3dDirac}) reduces to
		\begin{equation}
			\begin{aligned}
				0 &= \eta' + m(x) \eta - 2 p \chi\\
				0 &= \chi' - m(x) \chi 
			\end{aligned}\,, \label{eq:dwEom}
		\end{equation}
		where the $'$ denotes differentiation with respect to $x$. This solution must have definite helicity, with $\chi(x) = 0$, so that solutions take the form
		\begin{equation}
			\eta(x) = \mathcal{A} \es \exp\!\left(- \!\int_0^x\!\ud x' \, m(x')\right)\,, \label{eq:dwfProfile}
		\end{equation}
		where the coefficient $\mathcal{A}$ is determined by a normalization condition analogous to (\ref{eq:normalizationCond}).
		
		For (\ref{eq:dwfProfile}) to yield a normalizable zero mode propagating in the $(+z)$-direction, the mass $m(x)$ must be positive as $x \to \infty$ and negative as $x \to \minus \infty$, and thus \emph{vanish} for some value of $x$. Likewise, if the mass is instead negative as $x \to \infty$ and positive as $x \to \minus \infty$, there is a normalizable zero mode propagating in the $(\minus z)$-direction. Thus, if there exists a \emph{domain wall} in which the fermion mass crosses through zero, there will be chiral fermionic zero modes---often called chiral domain wall fermions---localized and propagating along it.\footnote{This is a well-known fact which has been exploited~\cite{Kaplan:1992sg} to simulate chiral fermions on the lattice. Many of the phenomena we find in this paper have analogs there. For instance, these chiral modes only exist for momenta $\mb{p}$ in a particular region of the Brillouin zone~\cite{Jansen:1992tw}, which we can denote as $\mb{p} \in \mathcal{C}$. For $\mb{p} \notin \mathcal{C}$, the chiral modes cease to exist. In analogy with Figure~\ref{fig:massiveZeroModePlot}, the chiral modes have wavefunctions that are well-localized to the domain wall for momenta comfortably inside $\mathcal{C}$, while they completely delocalize from the domain wall as $\mb{p}$ approaches the boundary $\partial \mathcal{C}$, and are no longer normalizable for $\mb{p} \notin \mathcal{C}$.} 
        We see that in $\mathbb{R}^{2,1}$, the fermion cannot be everywhere gapped and still yield a normalizable zero mode.
        This is one case in which vanishing mass and the existence of fermionic zero modes are inextricably linked.

		However, the situation changes if we restrict to the half-space $\mathbb{R}^{2,1}_\subp$, with $x \geq 0$. In this case, constant mass $m(x) = \mu$ \emph{does} yield a normalizable zero mode which is exponentially localized to the boundary at $x = 0$. For example, by decoupling (\ref{eq:dwEom}) we find that $\eta$ obeys
		\begin{equation}
			\eta'' + \big(m'(x) - m^2(x)\big)\eta = 0\,. \label{eq:etaEom}
		\end{equation}
		To match onto the analysis of \S\ref{sec:fzm} and \S\ref{sec:pert}, we note that, since $m(x)$ approaches a constant $\mu$ at infinity, then $\eta(x) \sim C_\subm \e^{\sminus \mu x} + C_\subp \e^{\es \mu x}$ as $x \to \infty$. Likewise, as long as $m(x)$ is regular as $x \to 0$, both linearly-independent solutions will be regular, and so we will always be able to construct a normalizable solution to (\ref{eq:etaEom}). When $m(x) = \mu$ is constant everywhere on $\mathbb{R}^{2, 1}_{\subp}$, this solution is just $\eta(x) = \mathcal{C}_\subm \e^{\sminus \mu x}$. In contrast, this fails on the full-space $\mathbb{R}^{2,1}$ because the solution that exponentially decays as $x \to \infty$ never matches onto the one which decays as $x \to \minus \infty$, unless the mass $m(x)$ switches sign for some $x$. We note in passing that the existence of this edge mode in the half-plane depends also on the choice of boundary conditions for the fermion; see e.g.,~\cite{Witten:2019bou}.

        To connect this simple system to the one we are interested in (\ref{eq:zmEomMass}), we can picture the polar radial coordinate in $\mathbb{R}^{3,1}$ as the analog of the $x$ coordinate on the half-space $\mathbb{R}^{2,1}_\subp$, while the axion string itself provides an effective ``mass'' $m(r)$ which is regular as $r \to 0$, as are the solutions to (\ref{eq:zmEomMass}). The same mechanism is at work for both cases: both linearly-independent solutions are regular at the core of the string or wall, and thus the solution that decays at spatial infinity is necessarily normalizable. As we discuss in the next section, this simple system is also useful for understanding the critical case in which core and Yukawa masses are equal, $m = \mu$, and the zero modes become completely delocalized from the string.

	\subsection{The Critical Mass Case} \label{sec:critMass}
		
		From the numerical results presented in Figure~\ref{fig:massiveZeroModePlot}, we found that there is a
        phase transition that occurs as we tune the core mass to the Yukawa mass, $m \to \mu$, wherein the fermionic zero mode seemingly delocalizes from the axion string. Beyond this critical mass, $m > \mu$, the zero mode ceases to exist. Since this critical case turns out to be very difficult to study numerically, it will instead be helpful to attack this case analytically to understand exactly how these modes behave when $m = \mu$. 

        First, however, it will be useful to qualitatively understand how solutions to  (\ref{eq:zmEomMass}) should behave in the limit $m \to \mu$. In the previous section, we described how a fermionic zero mode can arise whenever a fermion's spatially-dependent mass crosses through zero. At criticality and far from the string, the mass (\ref{eq:massMass}) approaches
		\begin{equation}
			\lim_{r \to \infty} M(r, \varphi) = \sqrt{(m - \mu \cos \varphi)^2 + \mu^2 \sin^2 \varphi}\,, \label{eq:asympMass}
		\end{equation}
		which crosses through zero at $\varphi =0$ when (and only when) $m = \mu$. From the logic of the previous section, we should then expect that the zero modes are no longer localized along the axion string, but are instead allowed to propagate freely along---and are localized to---the plane defined by $\varphi = 0$. Thus, we expect that these zero modes may have momentum along both the $x$- and $z$-directions, which we denote $p_x$ and $p_z$, respectively, with frequency $\omega = \sqrt{p_x^2 + p_z^2}$.

        We can exhibit these solutions in the critical case by searching for solutions to the equation of motion (\ref{eq:zmEomMass}) far from the string along the positive $x$-axis. In this limit, the equations of motion become approximately translationally invariant in both the $x$- and $z$-directions, and so we may search for solutions of the form
		\begin{equation}
			\psi(x) = \psi(x, y) \es\es \e^{ \sminus i \omega t + i p_x x + i p_z z}\,,
		\end{equation} 
		where $\psi(x, y)$ varies slowly along the $x$-direction.
        Since these modes must have definite helicity, we will assume an ansatz of the form 
        \begin{equation}
            \psi(x, y) \approx \sqrt{\frac{\omega}{2}}\begin{pmatrix} \cos \alpha \\ \sin \alpha \\ i \sin \alpha \\ \minus i \cos \alpha \end{pmatrix} F(x, y) \label{eq:criticalAnsatz}
        \end{equation}
        with $\cos 2\alpha = \minus p_z/\omega$, and  $\sin 2\alpha = \minus p_x/\omega$. With this ansatz, when far from the string (\ref{eq:zmEomMass}) reduces to
        \begin{equation}
           \partial_y F(x, y) \sim - \mu (y/x) F(x, y)\,,\mathrlap{\qquad x \gg |y|\,,}
        \end{equation} 
        where we have dropped derivatives with respect to $x$ and approximated $\e^{i \varphi} \approx 1 + i (y/x)$. This can be solved to find 
        \begin{equation}
            F(x, y) \approx \mathcal{A}(x) \, \exp\!\es\left(-\!\es\int_0^{|y|}\!\ud y'\, M(x, y)\right) \sim \left(\frac{\mu}{\pi x}\right)^{\frac{1}{4}} \exp\!\es\left(-\frac{\mu y^2}{2 x}\right), \label{eq:critMassSol03}
        \end{equation} 
        for $x \gg |y|$, where $M(x, y) = \mu \sqrt{(1 - \cos \varphi)^2 + \sin^2 \varphi} \approx \mu (y/x)$ is the mass (\ref{eq:asympMass}) far from the string, where $\mu x \gg 1$. The overall amplitude $\mathcal{A}(x)$ is determined by imposing the normalization condition,
        \begin{equation}
            \int\!\ud^3 x\, \psi_{\mb{p}}^\dagger(x) \psi_{\mb{p}'}(x) = (2 \pi)^2 \omega \, \delta^{(2)}(\mb{p} - \mb{p}')
        \end{equation}
        at any constant $t$, the direct analog of (\ref{eq:normalizationCond}). For the ansatz (\ref{eq:criticalAnsatz}), this translates into the requirement that $\int \!\ud y\, |F(x, y)|^2 = 1$.

        \begin{figure}
            \centering
            \includegraphics{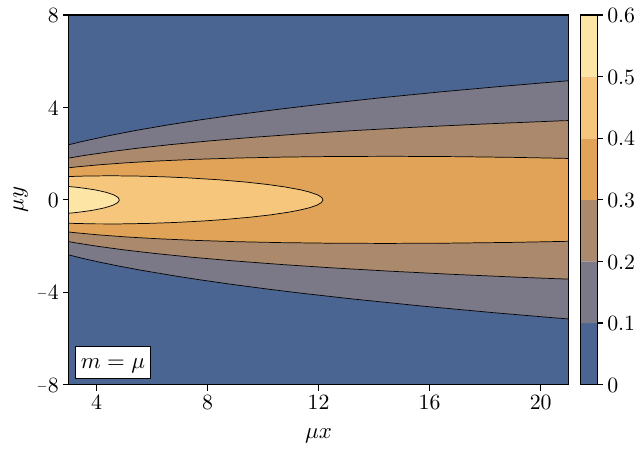}
            \caption{Plot of the approximate critical wavefunction $|F(x, y)|$, trustable for large positive $x$, with contours displayed in units of~$\sqrt{\mu}$. When $p_x = 0$ and $p_z = \minus \omega$, $|F(x, y)| = |\psi_0(x, y)|$, and so this figure can be directly compared to the bottom left panel of Figure~\ref{fig:massiveZeroModePlot}. \label{fig:critPlot}}
        \end{figure}

        We plot this approximate critical wavefunction (\ref{eq:critMassSol03}) in Figure~\ref{fig:critPlot} and find that it matches with the $m \to \mu$ behavior we observe in the numerical results shown in Figure~\ref{fig:massiveZeroModePlot}. As our qualitative arguments suggested, we find that the zero modes delocalize from the string once $m = \mu$ and instead are free to move along the half-plane defined by $\varphi = 0$. They are relatively well-localized along $y = \varphi = 0$, but become more and more spread out the further we get from the string, with an approximate width of $\Delta y \sim \sqrt{x}/\mu$. 

        To summarize, in this section we studied how fermionic zero modes along an axion string respond to the addition of a non-zero Dirac mass $m$. Absent this core mass, the four-dimensional fermion sees effectively zero mass at the core of the string, and so one might expect that this vanishing mass explains why the axion string supports fermionic zero modes. However, we showed that these zero modes exist even when the four-dimensional fermion has a core mass $m$ and is thus everywhere massive, and we explained what it is about the string that allows these modes to exist. Furthermore, we found that these zero modes completely delocalize from the string as $m \to \mu$, and cease to exist for $m > \mu$. In the next section, we explain this phase structure from the perspective of anomaly inflow, in which these zero modes are necessary to render the low-energy effective theory consistent.

\section{Zero Modes from Anomaly Inflow} \label{sec:inflow}

	The existence of these zero modes can also be inferred on topological grounds, based on the logic of anomaly inflow~\cite{Callan:1984sa}.
	Under an infinitesimal $\lab{U}(1)$ gauge transformation, the fields in (\ref{eq:massLag}) transform as
	\begin{equation}
	\delta_{\Lambda} \psi = i e \Lambda(x) \psi\, , \qquad
	\delta_{\Lambda} A_{\mu} = \partial_{\mu} \Lambda(x)\,,
	\label{eq:abelianGaugeTrans}
	\end{equation}
	where $\Lambda(x)$ is the $\lab{U}(1)$ gauge parameter. In a topologically trivial background, the action (\ref{eq:massLag}) is manifestly invariant under such a transformation. In the presence of the axion string, however, gauge invariance is more subtle. This symmetry is \emph{anomalous} in the presence of the string, such that electric charge is not conserved in a region localized to the string. The theory is then inconsistent unless there are some anomalous degrees of freedom, i.e. charged chiral excitations, that are localized to the string and can cancel this anomaly and carry away electric charge. These zero modes then allow electric charge and the anomaly to ``flow'' out of the bulk spacetime and onto the string, ensuring that the full theory remains consistent.

	The general strategy is as follows. We will attempt to construct a low-energy effective field theory of the axion and abelian gauge field by integrating out both the fermion $\psi(x)$ and radial mode $f(x)$ of the complex scalar. 
	We do this by first \emph{ignoring} the contribution from possible fermionic zero modes and by carefully considering the effect of a gauge transformation on the effective theory. Our arguments are similar to the original story, which can be found in \cite{Callan:1984sa,Naculich:1987ci,Harvey:2000yg}, except that the $\lab{U}(1)_{\slab{pq}}$ breaking mass leads to a subtlety in the identification of the phase which should be rotated away. Depending on the relative size of $m$ and $\mu$, we find a gauge anomaly localized to the string. This inconsistency then forces us to include an additional contribution from fermionic zero modes localized to the string which cancels this anomaly.

	We are interested in the gauge invariance of the low-energy effective theory for the axion field after integrating out the fermions. In the string background~(\ref{eq:axStringSol}), the fermion sector of the theory can be written
	\begin{equation}
	\mathcal{L} \supset \bar{\psi}\big( i\slashed{D} - m + y f(r) \e^{i\gamma^5 \theta(x)}\big) \psi, 
	\end{equation}
    with $\theta(x)$ the axion field, and so we wish to compute
    \begin{equation}
    \mathcal{Z}_{\psi}(\theta) = \int \! \mathcal{D}\bar{\psi} \mathcal{D}\psi\, \exp\!\bigg[\es\es i\!\int\! \dd^4 x\, \bar{\psi} \big(i\slashed{D} - M(r, \theta) \es\es \e^{i \gamma^5 \alpha(r, \theta)}\big)\psi \bigg]\,,
    \end{equation}
    with $M(r, \theta)$ and $\alpha(r, \theta)$ defined in (\ref{eq:massMass}).
	Performing the path integral by introducing a Pauli-Villars regulator with mass $\tilde{M}$, we find
    \begin{equation}
        \mathcal{Z}_{\psi}(\theta) = \frac{\det(i \slashed{D} - M(r, \theta)\e^{i\gamma^5\alpha(r, \theta)})}{\det(i\slashed{D} - \tilde{M})} \, .
    \end{equation}
	Now we perform a spatially-dependent chiral field redefinition, $\psi \to \e^{\sminus i\gamma^5 \alpha(r, \theta) /2} \psi$ to try and remove $\mathcal{Z}_\psi(\theta)$'s dependence on $\alpha(r, \theta)$. However, due to the chiral anomaly, this transformation introduces a Jacobian factor and we thus have
	\begin{equation}
	\label{eq:axion_eft}
	\mathcal{Z}_{\psi}(\theta) = \frac{\det(i\slashed{D} - M(r,\theta))}{\det(i\slashed{D}-\tilde{M})} \es \exp \bigg[ \frac{i}{8\pi^2}\! \int \! \alpha(r,\theta) F \wedge F \bigg],
	\end{equation}
	which can be written in the form of an effective action for the axion $\theta$ interacting with the gauge field $A_\mu$. 
    Note that in the $m \to 0$ limit, in which there is a classical $U(1)_{\textsc{pq}}$ symmetry, $\alpha(r, \theta) \to \theta$ and this Jacobian factor reduces to the usual quantized coupling of the axion to the gauge field, $\propto \theta F \wedge F$. Furthermore,
    integrating out the fermions leads to a nontrivial effective potential for the axion via the $\theta$-dependence in $M(r,\theta)$.

    \begin{figure}
            \centering
            \includegraphics{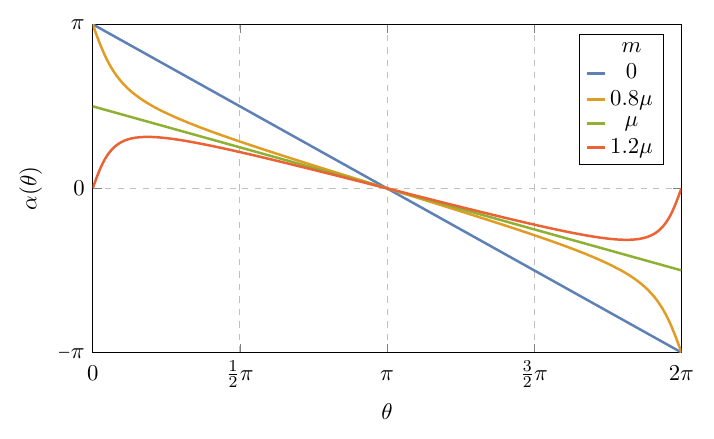}
            \caption{The phase of fermion ``mass'' (\ref{eq:phaseFar}) far from the string. The phase is double-valued at $\theta= 2\pi \sim 0$ for $m \leq \mu$ [{\color{Mathematica1} blue}, {\color{Mathematica2} orange}, {\color{Mathematica3} green}].  The zero mode solutions vanish as $m>\mu$ [{\color{Mathematica4} red}] once the phase becomes single-valued.}
            \label{fig:argMass}
    \end{figure}

	It is this additional $\alpha(r,\theta) F \wedge F$ term in the effective action that leads to a potential issue with gauge invariance. Whether or not this term is gauge invariant is determined by whether the phase far from the string,
	\begin{equation}
	\alpha(\theta) = \lim_{r \to \infty} \alpha(r, \theta) = \arg(m - \mu\es \e^{\sminus i\theta}) \,, \label{eq:phaseFar}
	\end{equation}
	is single- or multi-valued as a function of $\theta$, and this depends on the relative size of $m$ and $\mu$ as illustrated in Figure~\ref{fig:argMass}. For instance, in the limit that $m \ll \mu$, we find
	\begin{equation}
	\alpha(\theta) \approx \pi - \theta,
	\end{equation}
	which is clearly not single-valued at $\theta = 2\pi \sim 0$, and this applies more generally for all $m < \mu$. In this case, we must integrate the effective action by parts to make it well-defined,
	\begin{equation}
	\frac{1}{8\pi^2} \!\int \!\alpha(\theta) F \wedge F  \to -\frac{1}{8\pi^2} \! \int \!\dd \alpha(\theta) \wedge A \wedge F \, .
	\label{eq:delta_effective_action}
	\end{equation}
	However, under the gauge transformation (\ref{eq:abelianGaugeTrans}), this transforms as
	\begin{equation}
	\delta_{\Lambda} \bigg[ -\frac{1}{8\pi^2} \!\int \!\dd\alpha(\theta) \wedge A \wedge F \bigg] = -\frac{1}{8\pi^2} \!\int\! \dd\alpha(\theta) \wedge \dd(\es\Lambda F\es) = +\frac{1}{8\pi^2} \!\int \!\dd^2 \alpha(\theta) \wedge (\es\Lambda F\es) \,,
	\end{equation}
	where we have integrated by parts again in the last equality, neglecting any potential surface terms at the string core. Since $\alpha(\theta)$ is not single-valued, $\dd^2\alpha(\theta) \neq 0$ but is instead localized along the string, and we find an apparent gauge anomaly. 

    The reason for this apparent anomaly is that we have heretofore incorrectly assumed that there are no zero modes at the core of the string. Their existence modifies the chiral transformation in (\ref{eq:axion_eft}) and, with a careful treatment of the variation at the core,\footnote{
		Somewhat famously, arriving at the exact value of the gauge anomaly, which matches the contribution from two-dimensional chiral fermions localized on the string worldsheet, requires some care. A proper treatment involves introducing a ``bump-form'' which extends the validity of the effective theory in (\ref{eq:axion_eft}) to all of spacetime and properly includes the zero modes on the string core so that all surface contributions vanish~\cite{Harvey:2000yg}. This bump form can be computed explicitly in terms of the zero modes profiles, and the calculation of \cite{Harvey:2000yg} is unchanged in the case $m \neq 0$.}
	we find a violation of gauge invariance precisely equal and opposite to the gauge anomaly due to a massless chiral fermion localized to the string worldsheet,
	\begin{equation}
	\delta_{\Lambda}\bigg[ -\frac{1}{8\pi^2} \!\int \!\dd\alpha(\theta) \wedge A \wedge F \bigg] = \frac{1}{4\pi} \Lambda\, F\,.
	\end{equation}
	On the other hand, it is precisely when $m > \mu$ that the phase $\alpha(\theta)$ becomes single-valued since, near the end points at $\theta = 0$ and $2\pi$, we have $\alpha(0) = \alpha(2\pi) = \arg(m-\mu) = 0$. In this case the interaction $\propto \alpha(\theta)\, F \wedge F$ is both well-defined and single-valued, and gauge invariance is maintained without the need for additional degrees of freedom along the string. This analysis breaks down as $m \to \mu$ because it no longer makes sense to define an effective field theory for the axion and gauge field ``far'' from the string since, as evident from our numerical results in Figure~\ref{fig:massiveZeroModePlot} and discussed in \S\ref{sec:critMass}, the fermionic zero modes are no longer localized at small~$r$.

    It is worth noting that, from the IR perspective, these anomaly inflow arguments do not entirely determine the theory on the string. While we have focussed throughout on a particular UV model, in which the zero modes are unambiguously identified with chiral modes of the fermion, in general, the anomaly could be cancelled by a different theory on the string worldsheet. For strings with an even winding number, for instance, the anomaly can be cancelled by compact bosons living on the worldsheet. Anomaly inflow guarantees that the IR theory has a $\lab{U}(1)$ symmetry with a particular anomaly, while the precise theory on the string depends on the details of the model in the UV.

\begin{figure}[t]
    \centering
        \begin{tikzpicture}[very thick]
            \def\circSize{0.13}
            \draw[decoration={markings, mark=at position 0.3 with {\arrow[scale=-1.25, rotate=6]{stealth}}, mark=at position 0.6 with {\arrow[scale=-1.25, rotate=6]{stealth}}, mark=at position 0.98 with {\arrow[scale=-1.25, rotate=7]{stealth}}}, postaction=decorate] (0, 0) circle (0.9);
            \draw[densely dashed] (0, 0) ++ (45:0.9) -- (45:1.85) node[right, shift={(-0.05, 0.1)}] {$\partial_\nu \theta(x_0)$};
            \draw[decoration={snake, amplitude=0.8mm, segment length=2.5mm}, decorate] (0, 0) ++(-45:0.9) -- (-45:1.8);
            
            \begin{scope}[shift={(-45:1.8+\circSize)}]
                \draw[fill=white] (0, 0) circle (\circSize);
                \draw[line width=0.25mm, rotate=45] (-\circSize, 0) -- (\circSize, 0);
                \draw[line width=0.25mm, rotate=45] (0, -\circSize) -- (0, \circSize);
                \draw (\circSize, 0) node[right, shift={(-0.05, -0.1)}] {$A_\lambda(y)$};
            \end{scope}

            \begin{scope}[shift={(-180:0.9)}]
                \def\circSize{0.15}
                \draw[line width=0.25mm, rotate=45] (-\circSize, 0) -- (\circSize, 0);
                \draw[line width=0.25mm, rotate=45] (0, -\circSize) -- (0, \circSize);
                \draw (-\circSize, 0) node[left, shift={(0.05, 0)}] {$J_\mu(z)$};
            \end{scope}

        \end{tikzpicture}
    \caption{Leading order diagram for the current $\langle J^\mu(z)\rangle$ produced by a spatial variation of the axion $\partial_\nu \theta(x_0)$ in the presence of a background electromagnetic field $A_\lambda$, used by Goldstone and Wilczek to compute the effective action.}
    \label{fig:gw_loop}
    \end{figure}
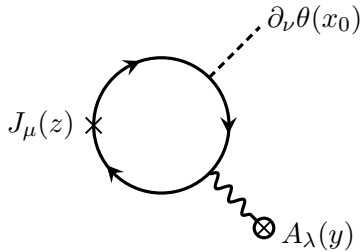

	Finally, it is illuminating to compare our effective action to the effective action computed via the method of Goldstone and Wilczek~\cite{Goldstone:1981kk}, as in~\cite{Callan:1984sa,Naculich:1987ci}. In that case, one computes the current at a point $x_0$ far from the string core, in response to a background electromagnetic field, in the long wavelength limit. 
    We choose $\theta(x_0) = 0$, so that the fermion-scalar interaction can be expanded as
    \begin{equation}
    y \bar{\psi} f(r) e^{i\gamma_5 \theta(x)} \psi \simeq i \mu\,\partial_{\lambda} \theta(x_0) (w - x_0)^{\lambda} \bar{\psi}\gamma^5 \psi .
    \end{equation}
    The leading contribution comes from the diagram in Figure~\ref{fig:gw_loop}. For the theory (\ref{eq:massLag}), this yields
    \begin{equation}
    \begin{aligned}
        \langle J^{\mu}(z) \rangle  = e^2 \mu\, \partial_{\lambda}\theta(x_0) \! \int \frac{\dd^4 p}{(2\pi)^4}& \frac{\dd^4 q}{(2\pi)^4} \frac{\dd^4 k}{(2\pi)^4}\, \dd^4 y\, \dd^4 w\, (w- x_0)^{\lambda} A_{\nu}(y)   \\[0.5em]
        &  \times \e^{\sminus i[p(z - y) + (p+q)(y-w) + (p+q+k)(w-z)]} \\ 
        & 
        \times \Tr\bigg[ \frac{1}{\slashed{p}-M}\gamma^{\nu} \frac{1}{\slashed{p}+\slashed{q}-M}\gamma^5 \frac{1}{\slashed{p}+\slashed{q}+\slashed{k}-M}\gamma^{\mu} \\
        & \qquad \quad+ \frac{1}{\slashed{p}-M}\gamma^{5} \frac{1}{\slashed{p}+\slashed{k}-M}\gamma^{\nu} \frac{1}{\slashed{p}+\slashed{k}+\slashed{q}-M}\gamma^{\mu}\bigg] 
    \end{aligned}
    \end{equation}
    and evaluates to
    \begin{equation}
        \langle J^{\mu}(z) \rangle  = \frac{e^2}{8\pi} \frac{\mu}{\mu - m} \epsilon^{\mu\nu\kappa\lambda} \partial_{\nu} \theta(x_0) F_{\kappa\lambda}\,.
    \label{eq:gw_current}
    \end{equation}
	Following \cite{Naculich:1987ci}, we can posit an effective action (valid far from the string),
    \begin{equation}
       S_\lab{eff} = \frac{1}{8 \pi^2} \frac{\mu}{\mu - m} \int\!\es \ud \theta \wedge A \wedge F\,,
        \label{eq:gw_effective_action}
    \end{equation}
	from which the current (\ref{eq:gw_current}) can be computed. When $m = 0$, integrating (\ref{eq:gw_effective_action}) by parts yields the usual, properly quantized, coupling of the axion to the gauge field. When $m \neq 0$, the $\mu / (\mu - m)$ prefactor would appear to violate the quantization condition for the axion--gauge field coupling. However, expanding for small values of the axion field, 
	\begin{equation}
	\dd \alpha(\theta) = \left[\frac{\mu}{\mu - m} + \mathcal{O}(\theta^2)\right] \dd \theta\,, 
	\end{equation}
	and we see that \eqref{eq:gw_effective_action} is precisely the leading term of (\ref{eq:delta_effective_action}), which is properly quantized. 

    Let us note that there is an ambiguity in which field one calls the ``axion,'' $\theta$ or $\alpha(\theta)$. Both are compact fields, $\theta \sim \theta + 2 \pi$ and $\alpha \sim \alpha + 2 \pi$. If one works in the ``charge-quantized'' basis in which the coefficient of the topological coupling (\ref{eq:delta_effective_action}) is an integer, the axion $\alpha$ has a (possibly highly) non-trivial metric on its field space given by 
    \begin{equation}
        g_{\alpha \alpha} = v^2 (\theta'(\alpha))^2 = v^2 \left[1 + \frac{m}{\mu} \frac{\cos \alpha}{\sqrt{1 - (m/\mu)^2 \sin^2 \alpha}}\right]^2 \,,
    \end{equation}
    and so around $\theta \approx 0$ or $\alpha \approx \pi$ the decay constant is effectively reduced by a factor of $(\mu - m)/\mu$, which is identical to the factor that appears in (\ref{eq:gw_effective_action}). If one instead works with $\theta$, the (classical) metric on field space is flat but the coefficient of the topological coupling (\ref{eq:gw_effective_action}) is not an integer.
    This story is similar in spirit to the reason for the non-quantized coupling of the axion to photons, due to mixing with the neutral pion.

	To summarize, we have shown that---with a careful treatment of the axion's periodicity---the arguments for anomaly inflow, and the resulting necessity of chiral zero modes on the string, persist in the presence of a field-independent mass, as long as this ``bulk'' mass is smaller than the mass generated by the chirally coupled scalar field. 
    This makes it clear that it is the topology of the axion string,
	and not the dynamics of the scalar field which spontaneously breaks the classical $\lab{U}(1)_{\slab{pq}}$ symmetry, which is responsible for the existence of the fermion zero modes.

\section{Low-Energy Effective Theory}
\label{sec:eft}

In the previous section, we found that the full four-dimensional effective theory was rendered consistent when $m < \mu$ by including chiral fermionic zero mode localized to the axion string, while they disappeared for $m> \mu$. Our numerical results in Figure~\ref{fig:massiveZeroModePlot} show that they disappear because the zero mode completely delocalizes from the string as $m \to \mu$. It will be useful to understand how this process appears from the zero mode's two-dimensional effective theory on the string worldsheet. At leading order in $\mu^{\sminus 1}$, this action is completely constrained by symmetry and does not depend on the ratio $m/\mu$. However, we will show that the Wilson coefficients of the higher derivative interactions between the zero mode and the gauge field depend sensitively on $m/\mu$ and diverge as $m \to \mu$, causing the two-dimensional effective theory to break~down.

We again will specialize to an axion string with charge $+1$, and start by introducing the (quantized) mode expansion for the fermion
\begin{equation}
\label{eq:zero_mode_operator}
\psi(x) = \int_0^{\infty}\! \frac{\dd p}{2\pi} \frac{1}{\sqrt{2|p|}}\,  \hat{a}_{p} \es \psi_{p}(r, \varphi)\es \e^{\sminus i p(t + z)} + \int_{-\infty}^0 \frac{\dd p}{2\pi} \frac{1}{\sqrt{2|p|}} \hat{b}_{\sminus p}^{\dagger} \es \psi_{p}(r, \varphi) \e^{\sminus i p(t+z)}+ \cdots
\end{equation}
where the $\cdots$ denote the non-zero modes of the massive four-dimensional fermion about the axion string. Here, $\psi_p(r, \varphi)$ is the zero mode wavefunction (\ref{eqn:zeromode}) found in Section~\ref{sec:massive}, subject to the normalization condition (\ref{eq:normalizationCond}). 
The creation and annihilation operators, $\hat{b}^{\dagger}_p$, $\hat{a}_p$ satisfy the canonical commutation relations appropriate for a {\em two-dimensional} fermion, 
\begin{equation}
\{ \hat{a}_{p}^{\vphantom{\dagger}}, \hat{a}_{p\smash{'}}^{\dagger}\} = 2\pi\, \delta(p - p'), \qquad
\{ \hat{b}^{\vphantom{\dagger}}_p, \hat{b}_{p\smash{'}}^{\dagger}\} = 2\pi\, \delta(p - p'). 
\end{equation}
Since the only place the momentum $p$ enters into $\psi_p(r, \varphi)$ is in its overall normalization, we can define the spinor $\mathcal{F}(r, \varphi) = \psi_p(r, \varphi)/\sqrt{p}$ and write (\ref{eq:zero_mode_operator}) as $\psi(x) = \chi_\sminus(t,z) \mathcal{F}(r, \varphi) + \cdots$, where 
\begin{equation}
    \chi_\sminus(t, z) = \int_{0}^{\infty} \!\frac{\ud p}{2 \pi} \, \Big[\hat{a}_p \es\es \e^{\sminus i p(t +z)} + \hat{b}^\dagger_{p}\es\es \e^{i p(t + z)}\Big]
\end{equation}
is a canonically normalized two-dimensional fermion field operator with negative chirality.

It will be convenient to repackage $\chi_\sminus(t, z)$ into a two-dimensional Dirac fermion $\chi(t, z)$ living on the axion string worldsheet 
\begin{equation}
    \chi(t, z) = \begin{pmatrix} \chi_\subm(t, z) \\ \chi_\subp(t, z) \end{pmatrix}\,,
\end{equation} 
by grouping it with a positive chirality fermion $\chi_\subp(t, z)$, which we later set to zero.
We will use $a, b, \ldots = 0, 3$ to denote worldsheet indices, and define our worldsheet $\gamma$-matrices as $\gamma^a = (\tilde{\gamma}^0, \tilde{\gamma}^3) = (\sigma_1, i \sigma_2)$\footnote{We use $\tilde{\gamma}^0$ and $\tilde{\gamma}^3$ to distinguish these two-dimensional $\gamma$-matrices from the four-dimensional $\gamma^0$ and $\gamma^3$, used for the four-dimensional Dirac fermion $\psi(x)$. } with chiral projectors $\gamma_\spm = \frac{1}{2}(\mathbbm{1} \pm \tilde{\gamma}^0 \tilde{\gamma}^3)$ such that $\gamma_\spm \chi = \chi_{\spm}$. We will thus impose the constraint $\gamma_\subp \chi = 0$ to remove $\chi_\subp$ from the theory. The kinetic term for the two-dimensional fermion then takes the standard Dirac form,
\begin{equation}
    S_\chi = \int\!\ud^2 \sigma\, i \bar{\chi} \gamma^a \partial_a \chi + \cdots = \int\!\ud^2 z\, i \chi^\dagger_\subm \partial_\subp \chi_\subm + \cdots
\end{equation}
where we denote $\partial_\pm = \partial_t \mp \partial_z$ and use $\ud^2 \sigma = \ud t \, \ud z$ to denote the volume element on the string worldsheet, while we will use $\ud^2 r = r \, \ud r \, \ud \varphi= \ud x\, \ud y$ to denote the volume element orthogonal to the string.  The $\cdots$ denote terms in the effective action that encode how the chiral zero mode $\chi_\subm$ interacts with the gauge field $A_\mu$, which we now derive.

The zero mode generates a four-dimensional current that points along the axion string
\begin{equation}
\label{eq:zero_mode_source}
    j^{\mu} =  e \bar{\psi} \gamma^{\mu}\psi = \begin{dcases} 
            e |\mathcal{F}(r, \varphi)|^2 (\bar{\chi}\gamma^{a}\chi)(t,z) & \mu = a=  0, 3 \\
            0 & \mu = 1, 2
        \end{dcases} \,,
\end{equation}
where $|\mathcal{F}(r, \varphi)|^2 = (\mathcal{F}^\dagger\mathcal{F})(r, \varphi)$, and so its interaction with the gauge field $A_\mu$ is determined by
\begin{equation}
    S_\slab{em} = - \!\int\!\ud^4 x\, j^\mu A_\mu = -\!\int\!\ud^4 x\, e |\mathcal{F}(r, \varphi)|^2 (\bar{\chi} \gamma^a \chi)(t, z) A_\mu(x)\,. \label{eq:zmGaugeInt}
\end{equation}
Our goal is to encode the information contained in (\ref{eq:zmGaugeInt}) into a series of effective interactions in the two-dimensional effective theory between $\chi$ and (derivatives of) the gauge field pulled back onto the string worldsheet. This will encode the zero mode wavefunction's multipolar structure into a set of effective two-dimensional interactions and will define a set of multipole moments for the current (\ref{eq:zero_mode_source}).

To derive these effective interactions, we perform a long-wavelength expansion of the gauge field about the axion string $x = (t, 0, 0, z)$ following a procedure similar to the one outlined in \cite{Ross:2012fc} and explained in more detail in Appendix~\ref{app:multipoles}. After using current conservation and integration by parts, (\ref{eq:zmGaugeInt}) can be written as
\begin{equation}
    \label{eq:multipole_action_unreduced}
    \begin{aligned}
        S_\slab{em} &= -\int\!\ud^2 \sigma  \big[e(\bar{\chi} \gamma^a \chi) A_a\big]\!\es (t, z)
        \\ 
        &- \sum_{n=1}^{\infty}\frac{1}{n!} \! \left[\int \! \dd^2 r\, x^{k_1} \cdots x^{k_n}\, |\mathcal{F}(r, \varphi)|^2 \right]  \!\int \!\ud^2 \sigma  \big[e (\bar{\chi} \gamma^a \chi)\partial_{k_2} \cdots \partial_{k_n} F_{k_1 a}\big]\!\es(t,z),
    \end{aligned}
\end{equation}
where we use $i, j, k_1, \cdots = 1, 2$ indices to denote directions orthgonal to the string, i.e.~${x^{i} = (x, y)}$. It is more illuminating to rewrite the higher order terms in this expansion in terms of irreducible $\lab{SO}(2)$ tensors. This introduces terms $\propto \partial^k F_{ka}$, which can be rewritten using the equations of motion that follow from \eqref{eq:axion_eft},
\begin{equation}
\partial^k F_{ka} = \minus\partial^{b}F_{ba} + \frac{e^2}{16\pi^2} \partial^{\mu} (\alpha(\theta) \tilde{F}_{a\mu}) + j_{a}\,,
\end{equation}
where $\tilde{F}_{\mu \nu} = \tfrac{1}{2} \epsilon_{\mu \nu \rho \sigma} F^{\rho \sigma}$ is the dual field strength. To second order in the derivative expansion, we find that the two-dimensional current $\tilde{\jmath}^{\es\es a} \equiv \bar{\chi} \gamma^a \chi$ couples to 
\begin{equation}
    \begin{aligned}
        S_\slab{em} \supset \int\!\ud^2 \sigma\!\left[-\tilde{\jmath}^{\es\es a} A_a + I_{1}^{i} \, \tilde{\jmath}^{\es\es a} F_{i a} + \frac{1}{4} I_{2}^{ij}  \tilde{\jmath}^{\es\es a} \partial_{(i} F_{j) a} - \frac{1}{4} I_2^\lab{tr} \tilde{\jmath}^{\es\es a} \partial^b F_{b a} + \frac{e^2}{64 \pi^2} I_2^\lab{tr} \tilde{\jmath}^{\es\es a} \partial^\rho(\theta \tilde{F}_{a \rho})\right]\label{eq:multipole_action}
    \end{aligned}
\end{equation}
where 
\begin{equation}
    I_{1}^i = \int \! \dd^2r \, x^i |\mathcal{F}(r, \varphi)|^2 \label{eq:dipVar}
\end{equation}
is the zero mode current's ``dipole,'' while 
\begin{equation}
    \begin{gathered}\label{eq:quadVar}
     \quad 
    I_{2}^{ij} = \int \! \dd^2 r\,  (x^i x^j - \tfrac{1}{2} \delta^{ij} x_k x^k) |\mathcal{F}(r, \varphi)|^2, \qquad I_{2}^\lab{tr} = \int\!\ud^2 r\, r^2 |\mathcal{F}(r, \varphi)|^2
    \end{gathered}\,,
\end{equation}
are the current's ``quadrupole'' and variance. Each of the operators in (\ref{eq:multipole_action}) are operators on the string worldsheet, while the dimensionless coefficients $c_1^i \equiv \mu I^i_1$, $c_2^{ij} \equiv \mu^2 I^{ij}_{2}$, $c_2^\lab{tr} \equiv \mu^2 I_2^\lab{tr}$ can be understood as Wilson coefficients in the effective theory on the string worldsheet.\footnote{There are only three independent Wilson coefficients that appear at each order in this multipole expansion. This can be most easily seen by remembering that the irreducible tensors of $\lab{SO}(2)$ with weight $\ell$ are in one-to-one correspondence with the two Fourier modes $\e^{\pm i \ell \varphi}$ and so, in general, correspond to two complex (or four real) degrees of freedom, which we may denote as $f_\ell$ and $f_{\sminus \ell}$. Since the distribution is purely real, these coefficients obey $f_\ell = f_{\sminus \ell}^*$, and so the distribution is described by only two real degrees of freedom at each weight $\ell$. To this, we must add the pure trace term at each $\ell$, e.g. $I_{2}^\lab{tr}$, which increases the number of moments at each order to three.} 

\begin{figure}[t]
\centering
\includegraphics[]{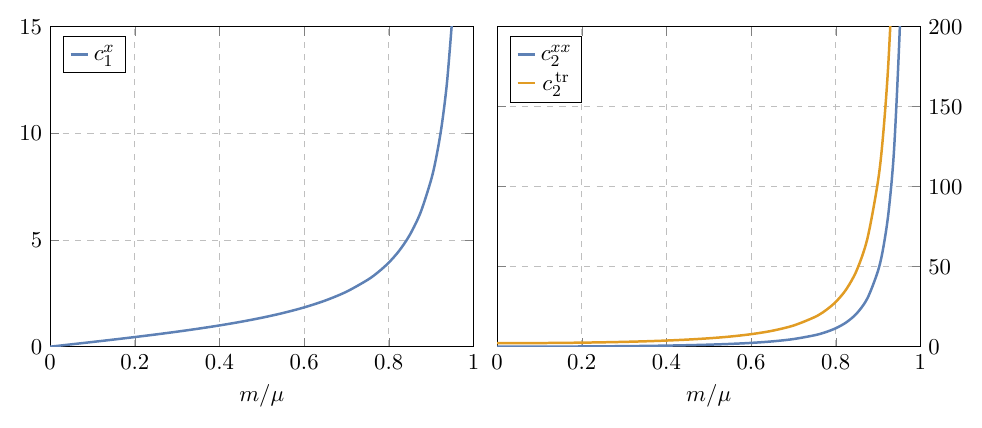}
\caption{Plots of the Wilson coefficients $c_{1}^x$, $c_{2}^{xx}$, and $c_2^\lab{tr}$ vs. $m / \mu$. The coefficients $c_1^y$ and $c_2^{xy} = c_2^{yx}$ vanish due to the zero mode's reflection symmetry about the $x$-axis, as evident in Figure~\ref{fig:massiveZeroModePlot}, while the other coefficient $c_2^{yy}$ is completely determined by $c_2^{yy} = \minus c_{2}^{xx}$. }
\label{fig:multipole_coeffs}
\end{figure}

We plot these Wilson coefficients in Figure~\ref{fig:multipole_coeffs} as functions of $m/\mu$. 
Two features are immediately obvious: at $m = 0$, the dipole and quadrupole moments vanish due to the cylindrical symmetry of the zero mode solution, as expected. The trace term $c_{2}^{\lab{tr}}$ does not vanish but instead approaches a value $c_{2}^\lab{tr} \to 2.19$ as $m \to 0$. Such trace terms also appear in the generic multipole expansion of a point-like source~\cite{Ross:2012fc} where, as here, they multiply \emph{time} derivatives of the field strength and thus do not appear in the static multipole expansion.

As $m$ approaches the critical value, $m = \mu$, all three Wilson coefficients quickly blow up, reflecting the breakdown of the two-dimensional effective theory. This is also expected since, as seen in Figure~\ref{fig:massiveZeroModePlot}, the zero modes delocalize from the string as $m \to \mu$ and must completely disappear for $m > \mu$, and so the worldsheet effective theory must fail. As is usual, the effective theory signals its own demise through Wilson coefficients that become uncontrollably large.

So, far from a charge $+1$ axion string the theory (\ref{eq:massLag}) is well-described by the effective action
\begin{equation}
    \begin{aligned}
        S &= \int\!\ud^4 x \left[-\frac{1}{4} F_{\mu \nu} F^{\mu \nu} + v^2 (\partial \theta)^2 + \frac{\alpha(\theta)}{16 \pi^2}  F_{\mu \nu} \tilde{F}^{\mu \nu} + \cdots \right] \\
        & + \int\!\ud^2 \sigma \left[\bar{\chi} i \slashed{D} \chi + \frac{c_1^i}{\mu}\tilde{\jmath}^{\es\es a} F_{i a} + \frac{c_2^{ij}}{4 \mu^2} \tilde{\jmath}^{\es\es a} \partial_{(i} F_{j) a} - \frac{c_2^\lab{tr}}{4\mu^2}  \tilde{\jmath}^{\es\es a} \partial^b F_{ba} + \frac{e^2 c_2^\lab{tr}}{64 \pi^2 \mu^2} \tilde{\jmath}^{\es\es a} \partial^\rho(\theta \tilde{F}_{a \rho}) + \cdots\right]
    \end{aligned}\,.
\end{equation}
The first line describes the dynamics of the gauge field and axion, where the $\cdots$ denote interactions that are induced by, for instance, integrating out the scalar's radial mode $f$ and the non-zero mode fluctuations of the fermions and include a potential for the axion~$\theta$. The second line describes the chiral zero mode living on the string and its interactions with the gauge field and axion. Here, the $\cdots$ denote terms that are either subleading in the derivative expansion or are nonlinear in the current $\tilde{\jmath}^{\es\es a}$, the latter of which are also generated by the long-wavelength expansion described above and in Appendix~\ref{app:multipoles} and upon integrating out the heavy scalar and fermionic modes. 


\section{Conclusions}
\label{sec:conclusions}

    In this work, we revisited the superconductivity of string solutions in a simple model of axion electrodynamics in which the $\lab{U}(1)_{\textsc{pq}}$ symmetry is explicitly broken by a mass term for the fermions.
    When the PQ-breaking mass $m$ is smaller than the asymptotic mass $\mu$ acquired from the radial mode of the scalar whose phase is the axion, the fermionic zero modes responsible for superconductivity persist. We demonstrated the existence of these zero modes both by studying the asymptotic behavior of the equations of motion and by solving for their profiles numerically. We also demonstrated how their existence can be understood from anomaly inflow, with some modifications to the original arguments by Callan and Harvey.

    For $m > \mu$, the zero modes cease to exist. As $m$ approaches $\mu$, however, the zero modes exhibit an interesting critical behavior, in which they delocalize from the core of the string and propagate along a two-dimensional wedge.\footnote{Amusingly, this behavior is very similar to how light fermions behave in the presence of a magnetic monopole~\cite{Callan:1982ah}, but in reverse. There, the light fermionic modes become more and more delocalized from the monopole core as their mass \emph{decreases}, ultimately explaining why the $\theta$-angle becomes a redundant parameter for dyon physics once $m = 0$. Here, counterintuitively, the light fermionic modes delocalize as their mass \emph{increases}.}
    We further studied the effective theory of the zero modes on the worldsheet and their interactions with external gauge fields. As $m$ approaches its critical value, we demonstrated explicitly how this effective theory breaks down as a result of the zero modes delocalizing.

    While we have focused on a simple ``minimal'' model of axion superconductivity here, there are several aspects which may be phenomenologically relevant in more realistic theories that warrant further exploration.
    First, as mentioned in the introduction, DFSZ axion models have multiple distinct, topologically stable string configurations and at least one of these (the ``Type-C strings'' in \cite{Abe:2020ure}) has a scalar profile which does not restore electroweak symmetry at the string core. It is possible that the nonzero vacuum expectation value at the core in these configurations may play a similar role to the explicit PQ-breaking mass term we have considered, which helps explain how these solutions may still be superconducting (as anomaly inflow arguments suggest they must be). This should be studied in more detail.

    It is also worth recalling that in a high-temperature background, fermions acquire a Debye mass $\sim g T$ which has the same effect as the explicit mass $m$. In an expanding universe, if the initial temperature is large enough that the Debye mass is larger than the Yukawa mass, a phase transition to the superconducting case may actually occur. It may thus be of interest to study if the delocalization of the zero modes found in the simple model studied here can have consequences for the evolution of string networks in cosmology.

    Finally, we should emphasize that our numerical results were derived under the assumption that the fermion amplitude is small. It would be worthwhile to understand how the profiles of the zero modes change beyond the limit of small fluctuations. In the same vein, it would be enlightening to understand the higher-order source terms in the worldsheet effective action that we have dropped. We hope to return to some of these topics in the future.


\section*{Acknowledgments}

We are grateful to Clay Cordova, Nicolas Fernandez, Junwu Huang, Austin Joyce, Seth Koren, Ho Tat Lam, Cody Long, Rashmish Mishra, Matthew Reece, Shu-Heng Shao and Tanmay Vachaspati for useful conversations. 
We are also especially grateful to Matthew Reece for comments on the manuscript.
HB, KF, SH and JS are supported in part by the DOE grant \texttt{DE-SC0013607}. KF and SH are also supported in part by the Alfred P.~Sloan Foundation Grant No.~\texttt{G-2019-12504}. HB, KF and JS are also supported in part by NASA Grant \texttt{80NSSC20K0506}.

\appendix

\section{Numerical Techniques} \label{app:numerics}

We rely on Chebyshev interpolation to numerically solve the zero mode equations of motion~(\ref{eq:zmEomMassComp}). 
This method is one of several numerical approaches used for collocation: the discretization of partial differential equations (PDEs) to transform them into matrix equations. Here we give a brief introduction; more details can be found in \cite{Trefethen:2013ata, Berrut:2004bli}.

Chebyshev interpolation provides a method of approximating a smooth function~$f(\zeta)$ on the interval $\zeta \in [\minus 1, 1]$ using its values at a finite and discrete set of points. That is, given the function $f(\zeta)$ evaluated at the set of \emph{Cheybshev points},
\begin{equation}
    \zeta_k = \cos \left(\frac{\pi(2 k +1)}{2 (N+1)}\right)\,,\mathrlap{\qquad k = 0, \ldots, N}\,, \label{eq:chebyshevPoints}
\end{equation}
we may approximate it at \emph{any} point in $\zeta \in [\minus 1, 1]$ via the interpolant
\begin{equation}
    f_N(\zeta) = \sum_{k = 0}^{N} f(\zeta_k) p_k(\zeta) \label{eq:chebyshevInterpolant}
\end{equation}
where the functions $p_k(\zeta)$, defined such that $p_k(\zeta_n) = \delta_{k n}$, are the degree-$N$ Lagrange polynomials associated to the points (\ref{eq:chebyshevPoints}). Importantly, the Lagrange polynomials can be reliably evaluated away from the interpolation points using the \emph{second barycentric form}~\cite{Higham:2004num,Berrut:2004bli},
\begin{equation}
    p_k(\zeta) = \frac{w_k}{\zeta - \zeta_k}\,  \Bigg/ \sum_{n = 0}^{N} \frac{w_n}{\zeta - \zeta_n}\,,
\end{equation} 
where we introduce the \emph{Chebyshev weights}
\begin{equation}
    w_k = \sin \left(\frac{2 \pi k(N+2) + \pi}{2(N+1)}\right)\,,\mathrlap{\qquad k = 0, \ldots, N\,.}
\end{equation} 
The interpolant (\ref{eq:chebyshevInterpolant}) then provides a degree-$N$ polynomial approximation that exactly agrees with $f(\zeta)$ at the points $\zeta_k$ and has an error which decays exponentially in $N$ for all other points on the interval.

The interpolant (\ref{eq:chebyshevInterpolant}) allows us to represent the function $f(\zeta)$ as a $(N+1)$-component vector $\vec{f} = \{f(\zeta_k)\}$. We can similarly represent derivatives of this function working with the derivative matrices $p'_k(\zeta_n)$ and $p''_{k}(\zeta_n)$, defined by~\cite{Berrut:2004bli,Baltensperger:2000imp}
\begin{equation}
    \begin{aligned}
        p'_k(\zeta_n) &= \begin{dcases}
                            \frac{w_k/w_n}{\zeta_n - \zeta_k} & n \neq k \\
                            - \sum_{k \neq n} p'_k(\zeta_n) & n = k
                        \end{dcases}\,\,\,, \\
        p''_k(\zeta_n) &= \begin{dcases}
                            2 p'_k(\zeta_n) p'_n(\zeta_n) - \frac{2 p'_k(\zeta_n)}{\zeta_n - \zeta_k} & n \neq k \\
                            - \sum_{k \neq n} p''_k(\zeta_n) & n = k
                        \end{dcases}\,\,\,.
    \end{aligned} \label{eq:chebyshevDMat}
\end{equation}
Any differential equation for $f(\zeta)$ on the interval $\zeta \in [\minus 1, 1]$ can then be approximated by a matrix equation for the function values $\{f(\zeta_k)\}$ by substituting (\ref{eq:chebyshevInterpolant}) for $f(\zeta)$, evaluating the equation at each Chebshev point $\zeta_k$, and using (\ref{eq:chebyshevDMat}) to convert differentiation into matrix multiplication. This matrix equation can then be solved by any number of standard numerical techniques.

For example, the equation
\begin{equation}
    \left[\frac{\ud^2}{\ud \zeta^2} + g(\zeta)\right]\!f(\zeta) = 0\,, \label{eq:exampleDEQ}
\end{equation}
for any regular function $g(\zeta)$ on $\zeta \in [\minus 1, 1]$ can be approximated by the matrix equation
\begin{equation}
    \sum_{k = 0}^{N} \left(\mathcal{D}_2 + \mathcal{G}\right)_{nk} f(\zeta_k) = 0\,,
\end{equation}
where $(\mathcal{G})_{nk} = g(\zeta_k) \delta_{nk}$, $(\mathcal{D}_2)_{nk} = p_{k}''(\zeta_n)$, which is defined in (\ref{eq:chebyshevDMat}), and the number of nodes is chosen to be $N+1$. The solutions to (\ref{eq:exampleDEQ}) are then well-approximated by the interpolant (\ref{eq:chebyshevInterpolant}) constructed from the null space of the matrix $(\mathcal{D}_2 + \mathcal{G})$. As long as $N$ is large enough, this interpolant then provides an approximate solution to (\ref{eq:exampleDEQ}) whose pointwise error decays exponentially in $N$.

When the equation we aim to solve is defined on the semi-infinite line $r \in [0, \infty)$, we must first map it to the interval $\zeta \in [\minus 1, 1]$ to apply Chebyshev interpolation. We find that the simplest map, the linear fractional transformation
\begin{equation}
    \zeta(r) = \frac{r - L}{r+L}, \label{eq:MapOne}
\end{equation}
works well, where $L$ is a tuneable parameter that broadly controls which $r$ values are sampled by the corresponding Chebyshev nodes (\ref{eq:chebyshevPoints}). In practice, we take $L$ to be an $\mathcal{O}(1)$ number times the scale over which the solution varies naturally (set by $\mu^{\sminus 1}$ for the equations we are interested in) though we ensure that our solutions do not depend on the precise choice of $L$.  

For two-dimensional equations of motion like (\ref{eq:zmEomMassComp}), we have a choice in how to apply Chebyshev interpolation. 
The simplest approach is to expand in Fourier modes, as in \S\ref{sec:pert}, to reduce the problem to a system of coupled ordinary differential equations. That is, we expand each spinor component into their lowest $2J+1$ Fourier modes
\begin{equation}\label{eq:expandedPsi}
    \psi_{\alpha}(r, \varphi) = \sum_{|\ell| \leq J} \psi_{\alpha, \ell}(r)\, \e^{i \ell \varphi}\,,
\end{equation}
such that (\ref{eq:zmEomMassComp}) reduces to a finite set of coupled equations  (\ref{eq:zmEomMassFourier}) involving $2\times (2J+1)$ mode functions. We can then interpolate each radial function $\psi_{\alpha, \ell}(r)$ with $N+1$ nodes to convert (\ref{eq:zmEomMassFourier}) into a $2\times(2 J + 1) \times (N+1)$-dimensional linear equation.
The profiles displayed in Figure~\ref{fig:massiveZeroModePlot} were generated with $J = 41$ and $N = 61$, though our results are robust to changing both $J$ and~$N$.

Another approach is to interpolate the spinor components $\psi_{\alpha}(x, y)$ using a two-dimensional Cartesian grid. This requires first mapping both $x$ and $y$ from $(\minus \infty, \infty)$ to $[\minus 1, 1]$. We do so via two different maps
\begin{equation}\label{eq:MapsX}
    \zeta(x_i) = \begin{dcases}
         \arctan\es({x_i}/{L_{i}})& 0\leq m/\mu\leq 0.8 \\
        \tanh \es ({x_i}/{L_{i}}) & 0.4 \leq m/\mu \leq 0.95
    \end{dcases}\,,
\end{equation}
depending on the value of $m/\mu$, where $x_i = \{x, y\}$. 
Here, the parameters $L_{i} = \{L_x, L_y\}$ determine the relevant scales in the original $x$ or $y$ coordinate. These maps efficiently sample small $(x,y)$ values while avoiding numerical artifacts at large $x$ and $y$. Since the solutions elongate in the $x$-direction as $m \to \mu$, we used $L_x \simeq \mathcal{O}(1) \times 1/(\mu - m)$ and $L_y \simeq \mathcal{O}(1) \times \mu^{\sminus 1}$. We find that with these ranges of $L_x$ and $L_y$, our Cartesian results agree with each other for overlapping $m/\mu$ and that our solutions are independent of our specific choice. 
For large numbers of nodes, $N_x, N_y \geq 90$, our results obtained with a Cartesian grid agree with those obtained using a Fourier mode expansion, but formulating the problem in the Fourier basis is significantly faster.

Finally, to compute the moments of our solutions and properly normalize them, we must numerically evaluate integrals involving these profiles. We do so via quadrature, in which the integral is approximated by a weighted sum of the function's values on the Chebyshev grid. Explicitly, an arbitrary integral of the density $|\psi(r, \varphi)|^2 = |\psi_0(r, \varphi)|^2 + |\psi_3(r, \varphi)|^2$ can be expressed as
\begin{equation}
    \begin{aligned}\label{eq:quadratureInt}
        \int \!\ud^2 r \, a(r,\varphi) \es\es |\psi(r, \varphi)|^2 
        &= \sum_{\alpha,n, \ell, \ell'}{} \int_{\sminus 1}^{1}\!\ud \zeta \, \int_{0}^{2\pi}\! \ud \varphi\,  r'(\zeta) a_n(\zeta) \es \e^{in\varphi} \e^{i(\ell-\ell')\varphi}\es \psi^{\phantom{*}}_{\alpha,\ell}(\zeta)\es\psi^{*}_{\alpha, \ell'}(\zeta) \\
        & = 2\pi\!\sum_{\alpha, k,\ell, n} q_k\es\es r'(\zeta_k) \es a_n(\zeta_k)\es \psi^{\phantom{*}}_{\alpha, \ell}(\zeta_k)\es\psi^{*}_{\alpha, \ell+n}(\zeta_k)\,
     \end{aligned}
\end{equation}
where 
\begin{equation}
    q_k = \int_{\sminus 1}^{1}\!\ud \zeta\, p_k(\zeta)
\end{equation}
is the quadrature weight, which can be efficiently numerically evaluated, and
\begin{equation}
    a(r, \varphi) = \sum_{|n| \leq J, k} a_n(\zeta_k) p_k(\zeta) \e^{i n \varphi}
\end{equation}
is an arbitrary weight function, whose $n$'th Fourier mode has value $a_n(\zeta_k)$ at $r(\zeta_k)$. For instance, to normalize the profile functions in Figure~\ref{fig:massiveZeroModePlot},  we evaluate (\ref{eq:quadratureInt}) with $a(r, \varphi) = 1$, while to compute the dipole (\ref{eq:dipVar}) we choose $a(r, \varphi) \propto r \cos \varphi$ and $a(r, \varphi) \propto r \sin \varphi$.

\section{Multipole Moments}
\label{app:multipoles}

In this Appendix we provide more details on the derivation of the worldsheet effective action given in \eqref{eq:multipole_action_unreduced}. We follow the method in \cite{Ross:2012fc}, adapted to the case of a string. Our starting point is the term in the action that couples the gauge field to the four-dimensional current from \eqref{eq:zero_mode_source}, which is pointing along the axion string,
\begin{equation}
S_\slab{em} = - \!\int\!\ud^4 x\; j^\mu A_\mu .
\end{equation}
As in the main text, we take $\mu = 0,1,2,3$, $a = 0,3$, and $i, j, k_i = 1,2$.

Our goal is an effective action describing these interactions which is valid when the wavelength of the gauge field is much larger than the characteristic size of the source. In this regime, we can Taylor expand the gauge field about the string source at the origin,
\begin{equation}
A_\mu(t, \vec{x}) = - \! \sum_{n = 0}^{\infty} \frac{1}{n!}  \; x^{k_1} \! \cdots x^{k_n}\big[ \partial_{k_1} \! \cdots \partial_{k_n} A_\mu \big]_{x = (t, 0, 0, z)}.
\end{equation}
Since our source of interest points along the string (see \ref{eq:zero_mode_source}), we can drop the terms with $\mu = 1, 2$. We are left with the action
\begin{equation}
\label{eq:Sem_expanded_1}
S_\slab{em}= - \! \sum_{n = 0}^{\infty}  \int \!\ud^2 \sigma  \! \int \! \dd^2 r\,  \frac{1}{n!}  x^{k_1} \! \cdots x^{k_n}  j^{a}\!\es(t, \vec{x})\big[ \partial_{k_1} \! \cdots \partial_{k_n} A_a\big] \!\es(t,z) .
\end{equation}
It is useful to rewrite the action in terms of gauge invariant quantities. We first consider the $A^0$ term, and note that $E^{k_i} = F^{0 k_i} = \dot{A}^{k_i} + \partial_{k_i} A^0$.
This allows us to exchange radial derivatives of $A^0$ for time-derivatives of $A^{k_i}$ and terms involving the electric field, $E$. We find,
\begin{equation}
\label{eq:Sem_expanded_2}
\begin{split}
S_{A_0} = &- \int \!\ud^2 \sigma \int \! \ud^2 r \; j^0 \big[A^0\big]\!\es(t, z) - \sum_{n = 1}^\infty \frac{1}{n!} \int \!\ud^2 \sigma \int \! \dd^2 r \; j^0 x^{k_1} \!\cdots x^{k_n}\big[ \partial_{k_1} \! \cdots \partial_{k_{n-1}}E^{k_n}\big] \!\es(t,z) \\
&+ \sum_{n = 1}^\infty \frac{1}{n!}\int \!\ud^2 \sigma  \int \! \ud^2 r \; j^0 x^{k_1} \! \cdots x^{k_n}\big[\partial_{k_1} \!\cdots \partial_{k_{n-1}} \dot{A}^{k_n}\big] \!\es(t,z)
\end{split}
\end{equation}
The last term in this expression can be combined with the $A^3$ term of \eqref{eq:Sem_expanded_1} to get a gauge invariant term involving the magnetic field.
To see this, we integrate by parts within the $\dd^2\sigma$ integral, and use current conservation: $\partial_0 j^0 = - \partial_3 j^3$. This allows us to rewrite,
\begin{equation}
\begin{split}
\left[ \int \! \dd^2 r \; j^0 x^{k_1} \!\cdots x^{k_n} \right] \partial_{k_1} \!\cdots \partial_{k_{n-1}} \dot{A}^{k_{n}} 
& = 
\left[ \int \! \dd^2 r \; (\partial_3 j^3 )\;  x^{k_1} \!\cdots x^{k_n}\right ] \partial_{k_1} \!\cdots \partial_{k_{n-1}} A^{k_n} \\
& = 
-
\left[ \int \! \dd^2 r \; j^3 x^{k_1}\! \cdots x^{k_n}\right] \partial_{k_1}\! \cdots \partial_{k_{n-1}} \partial_3 A^{k_n},
\end{split}
\end{equation}
where in the last equality we have integrated by parts again. 
With this replacement, the sum of the last term of \eqref{eq:Sem_expanded_2} and the $A^3$ term from \eqref{eq:Sem_expanded_1} is
\begin{equation}
\begin{split}
S_{B} = &- \int \!\ud^2 \sigma \int \! \dd^2 r \; j^3 A_3 -  \sum_{n = 1}^{\infty} \frac{1}{n!} \int \!\ud^2 \sigma \int \! \dd^2 r \; j^3 x^{k_1} \! \cdots x^{k_n} \partial_{k_1} \! \cdots \partial_{k_{n-1}} \big(\partial_3 A^{k_n}\big) \\
&- \sum_{n = 1}^{\infty}  \frac{1}{n!} \int \!\ud^2 \sigma \int \! \dd^2 r \;  j^3  x^{k_1}\!  \cdots x^{k_n} \partial_{k_1} \! \cdots \partial{k_{n-1}} \big( \partial_{k_n} A_3 \big)\\
= &- \int \!\ud^2 \sigma \int \! \dd^2 r \;  j^3 A_3 -  \sum_{n = 1}^{\infty} \frac{1}{n!} \int \!\ud^2 \sigma \int \! \dd^2 r \;  j^3 x^{k_1} \! \cdots x^{k_n} \partial_{k_1} \! \cdots \partial_{k_{n-1}} \epsilon^{3 k_n j} B_j,
\end{split}
\end{equation}
where we have used $\partial_i A_3 - \partial_3 A_i = F^{i3} = \epsilon^{3ij} B_j$. Adding back the remaining terms in $S_{A^0}$ and relabeling indices, we obtain the full action
\begin{equation}
\begin{split}
S_\slab{em}  = &- \int \!\ud^2 \sigma \int \! \dd^2 r \; (j^0 A_0 + j^3 A_3) - \sum_{n = 1}^\infty \frac{1}{n!} \int \!\ud^2 \sigma \left[ \int \! \dd^2 r j^0 x^{k_1} \! \cdots x^{k_n}  \right] \partial_{k_2}\!  \cdots \partial_{k_n}E^{k_1} \\
&- \sum_{n = 1}^{\infty} \frac{1}{n!} \int \!\ud^2 \sigma \left[ \int \! \dd^2 r \; j^3 x^{k_1} \! \cdots x^{k_n}\right] \partial_{k_2} \! \cdots \partial_{k_n} \epsilon^{3 k_1 j} B_j,
\end{split}
\end{equation}
which can be written in terms of the field strength as
\begin{equation}
\begin{split}
S_\slab{em}  = &- \int \!\ud^2 \sigma \int \! \dd^2 r \; (j^0 A_0 + j^3 A_3) - \sum_{n = 1}^\infty \frac{1}{n!} \int \!\ud^2 \sigma  \left[ \int \! \dd^2 r j^a x^{k_1}\!  \cdots x^{k_n} \right] \partial_{k_2}\!  \cdots \partial_{k_n}F_{k_1 a}.
\end{split}
\end{equation}
This action reduces to \eqref{eq:multipole_action_unreduced} using the current \eqref{eq:zero_mode_source} and the normalization condition for $|\mathcal{F}(r, \varphi)|^2$.

\phantomsection
\addcontentsline{toc}{section}{References}
\bibliographystyle{utphys}
\bibliography{axionStrings.bib}

\providecommand{\href}[2]{#2}\begingroup\raggedright\begin{thebibliography}{10}

\bibitem{Peccei:1977ur}
R.~D. Peccei and H.~R. Quinn, ``{Constraints Imposed by CP Conservation in the
  Presence of Instantons},''
  \href{http://dx.doi.org/10.1103/PhysRevD.16.1791}{{\em Phys. Rev. D}
  {\bfseries 16} (1977) 1791--1797}.

\bibitem{Peccei:1977hh}
R.~D. Peccei and H.~R. Quinn, ``{CP Conservation in the Presence of
  Instantons},'' \href{http://dx.doi.org/10.1103/PhysRevLett.38.1440}{{\em
  Phys. Rev. Lett.} {\bfseries 38} (1977) 1440--1443}.

\bibitem{Weinberg:1977ma}
S.~Weinberg, ``{A New Light Boson?},''
  \href{http://dx.doi.org/10.1103/PhysRevLett.40.223}{{\em Phys. Rev. Lett.}
  {\bfseries 40} (1978) 223--226}.

\bibitem{Wilczek:1977pj}
F.~Wilczek, ``{Problem of Strong $P$ and $T$ Invariance in the Presence of
  Instantons},'' \href{http://dx.doi.org/10.1103/PhysRevLett.40.279}{{\em Phys.
  Rev. Lett.} {\bfseries 40} (1978) 279--282}.

\bibitem{Preskill:1982cy}
J.~Preskill, M.~B. Wise, and F.~Wilczek, ``{Cosmology of the Invisible
  Axion},'' \href{http://dx.doi.org/10.1016/0370-2693(83)90637-8}{{\em Phys.
  Lett. B} {\bfseries 120} (1983) 127--132}.

\bibitem{Dine:1982ah}
M.~Dine and W.~Fischler, ``{The Not So Harmless Axion},''
  \href{http://dx.doi.org/10.1016/0370-2693(83)90639-1}{{\em Phys. Lett. B}
  {\bfseries 120} (1983) 137--141}.

\bibitem{Abbott:1982af}
L.~F. Abbott and P.~Sikivie, ``{A Cosmological Bound on the Invisible Axion},''
  \href{http://dx.doi.org/10.1016/0370-2693(83)90638-X}{{\em Phys. Lett. B}
  {\bfseries 120} (1983) 133--136}.

\bibitem{Freese:1990rb}
K.~Freese, J.~A. Frieman, and A.~V. Olinto, ``{Natural inflation with pseudo -
  Nambu-Goldstone bosons},''
  \href{http://dx.doi.org/10.1103/PhysRevLett.65.3233}{{\em Phys. Rev. Lett.}
  {\bfseries 65} (1990) 3233--3236}.

\bibitem{Alexander:2004us}
S.~H.-S. Alexander, M.~E. Peskin, and M.~M. Sheikh-Jabbari, ``{Leptogenesis
  from gravity waves in models of inflation},''
  \href{http://dx.doi.org/10.1103/PhysRevLett.96.081301}{{\em Phys. Rev. Lett.}
  {\bfseries 96} (2006) 081301},
  \href{http://arxiv.org/abs/hep-th/0403069}{{\ttfamily arXiv:hep-th/0403069}}.

\bibitem{Witten:1984dg}
E.~Witten, ``{Some Properties of O(32) Superstrings},''
  \href{http://dx.doi.org/10.1016/0370-2693(84)90422-2}{{\em Phys. Lett. B}
  {\bfseries 149} (1984) 351--356}.

\bibitem{Svrcek:2006yi}
P.~Svrcek and E.~Witten, ``{Axions In String Theory},''
  \href{http://dx.doi.org/10.1088/1126-6708/2006/06/051}{{\em JHEP} {\bfseries
  06} (2006) 051}, \href{http://arxiv.org/abs/hep-th/0605206}{{\ttfamily
  arXiv:hep-th/0605206}}.

\bibitem{Arvanitaki:2009fg}
A.~Arvanitaki, S.~Dimopoulos, S.~Dubovsky, N.~Kaloper, and J.~March-Russell,
  ``{String Axiverse},''
  \href{http://dx.doi.org/10.1103/PhysRevD.81.123530}{{\em Phys. Rev. D}
  {\bfseries 81} (2010) 123530},
  \href{http://arxiv.org/abs/0905.4720}{{\ttfamily arXiv:0905.4720 [hep-th]}}.

\bibitem{Demirtas:2018akl}
M.~Demirtas, C.~Long, L.~McAllister, and M.~Stillman, ``{The Kreuzer-Skarke
  Axiverse},'' \href{http://dx.doi.org/10.1007/JHEP04(2020)138}{{\em JHEP}
  {\bfseries 04} (2020) 138}, \href{http://arxiv.org/abs/1808.01282}{{\ttfamily
  arXiv:1808.01282 [hep-th]}}.

\bibitem{Demirtas:2021gsq}
M.~Demirtas, N.~Gendler, C.~Long, L.~McAllister, and J.~Moritz, ``{PQ
  axiverse},'' \href{http://dx.doi.org/10.1007/JHEP06(2023)092}{{\em JHEP}
  {\bfseries 06} (2023) 092}, \href{http://arxiv.org/abs/2112.04503}{{\ttfamily
  arXiv:2112.04503 [hep-th]}}.

\bibitem{Stout:2020uaf}
J.~Stout, ``{Instanton expansions and phase transitions},''
  \href{http://dx.doi.org/10.1007/JHEP05(2022)168}{{\em JHEP} {\bfseries 05}
  (2022) 168}, \href{http://arxiv.org/abs/2012.11605}{{\ttfamily
  arXiv:2012.11605 [hep-th]}}.

\bibitem{Fan:2021ntg}
J.~Fan, K.~Fraser, M.~Reece, and J.~Stout, ``{Axion Mass from Magnetic Monopole
  Loops},'' \href{http://dx.doi.org/10.1103/PhysRevLett.127.131602}{{\em Phys.
  Rev. Lett.} {\bfseries 127} no.~13, (2021) 131602},
  \href{http://arxiv.org/abs/2105.09950}{{\ttfamily arXiv:2105.09950
  [hep-ph]}}.

\bibitem{Vilenkin:1982ks}
A.~Vilenkin and A.~E. Everett, ``{Cosmic Strings and Domain Walls in Models
  with Goldstone and PseudoGoldstone Bosons},''
  \href{http://dx.doi.org/10.1103/PhysRevLett.48.1867}{{\em Phys. Rev. Lett.}
  {\bfseries 48} (1982) 1867--1870}.

\bibitem{Sikivie:1982qv}
P.~Sikivie, ``{Of Axions, Domain Walls and the Early Universe},''
  \href{http://dx.doi.org/10.1103/PhysRevLett.48.1156}{{\em Phys. Rev. Lett.}
  {\bfseries 48} (1982) 1156--1159}.

\bibitem{Davis:1986xc}
R.~L. Davis, ``{Cosmic Axions from Cosmic Strings},''
  \href{http://dx.doi.org/10.1016/0370-2693(86)90300-X}{{\em Phys. Lett. B}
  {\bfseries 180} (1986) 225--230}.

\bibitem{Harari:1987ht}
D.~Harari and P.~Sikivie, ``{On the Evolution of Global Strings in the Early
  Universe},'' \href{http://dx.doi.org/10.1016/0370-2693(87)90032-3}{{\em Phys.
  Lett. B} {\bfseries 195} (1987) 361--365}.

\bibitem{Shellard:1987bv}
E.~P.~S. Shellard, ``{Cosmic String Interactions},''
  \href{http://dx.doi.org/10.1016/0550-3213(87)90290-2}{{\em Nucl. Phys. B}
  {\bfseries 283} (1987) 624--656}.

\bibitem{Davis:1989nj}
R.~L. Davis and E.~P.~S. Shellard, ``{Do Axions Need Inflation?},''
  \href{http://dx.doi.org/10.1016/0550-3213(89)90187-9}{{\em Nucl. Phys. B}
  {\bfseries 324} (1989) 167--186}.

\bibitem{Hagmann:1990mj}
C.~Hagmann and P.~Sikivie, ``{Computer simulations of the motion and decay of
  global strings},'' \href{http://dx.doi.org/10.1016/0550-3213(91)90243-Q}{{\em
  Nucl. Phys. B} {\bfseries 363} (1991) 247--280}.

\bibitem{Battye:1993jv}
R.~A. Battye and E.~P.~S. Shellard, ``{Global string radiation},''
  \href{http://dx.doi.org/10.1016/0550-3213(94)90573-8}{{\em Nucl. Phys. B}
  {\bfseries 423} (1994) 260--304},
  \href{http://arxiv.org/abs/astro-ph/9311017}{{\ttfamily
  arXiv:astro-ph/9311017}}.

\bibitem{Battye:1994au}
R.~A. Battye and E.~P.~S. Shellard, ``{Axion string constraints},''
  \href{http://dx.doi.org/10.1103/PhysRevLett.73.2954}{{\em Phys. Rev. Lett.}
  {\bfseries 73} (1994) 2954--2957},
  \href{http://arxiv.org/abs/astro-ph/9403018}{{\ttfamily
  arXiv:astro-ph/9403018}}. [Erratum: Phys.Rev.Lett. 76, 2203--2204 (1996)].

\bibitem{Yamaguchi:1998gx}
M.~Yamaguchi, M.~Kawasaki, and J.~Yokoyama, ``{Evolution of axionic strings and
  spectrum of axions radiated from them},''
  \href{http://dx.doi.org/10.1103/PhysRevLett.82.4578}{{\em Phys. Rev. Lett.}
  {\bfseries 82} (1999) 4578--4581},
  \href{http://arxiv.org/abs/hep-ph/9811311}{{\ttfamily arXiv:hep-ph/9811311}}.

\bibitem{Klaer:2017ond}
V.~B.~. Klaer and G.~D. Moore, ``{The dark-matter axion mass},''
  \href{http://dx.doi.org/10.1088/1475-7516/2017/11/049}{{\em JCAP} {\bfseries
  11} (2017) 049}, \href{http://arxiv.org/abs/1708.07521}{{\ttfamily
  arXiv:1708.07521 [hep-ph]}}.

\bibitem{Gorghetto:2018myk}
M.~Gorghetto, E.~Hardy, and G.~Villadoro, ``{Axions from Strings: the
  Attractive Solution},'' \href{http://dx.doi.org/10.1007/JHEP07(2018)151}{{\em
  JHEP} {\bfseries 07} (2018) 151},
  \href{http://arxiv.org/abs/1806.04677}{{\ttfamily arXiv:1806.04677
  [hep-ph]}}.

\bibitem{Vaquero:2018tib}
A.~Vaquero, J.~Redondo, and J.~Stadler, ``{Early seeds of axion
  miniclusters},'' \href{http://dx.doi.org/10.1088/1475-7516/2019/04/012}{{\em
  JCAP} {\bfseries 04} (2019) 012},
  \href{http://arxiv.org/abs/1809.09241}{{\ttfamily arXiv:1809.09241
  [astro-ph.CO]}}.

\bibitem{Buschmann:2019icd}
M.~Buschmann, J.~W. Foster, and B.~R. Safdi, ``{Early-Universe Simulations of
  the Cosmological Axion},''
  \href{http://dx.doi.org/10.1103/PhysRevLett.124.161103}{{\em Phys. Rev.
  Lett.} {\bfseries 124} no.~16, (2020) 161103},
  \href{http://arxiv.org/abs/1906.00967}{{\ttfamily arXiv:1906.00967
  [astro-ph.CO]}}.

\bibitem{Gorghetto:2020qws}
M.~Gorghetto, E.~Hardy, and G.~Villadoro, ``{More axions from strings},''
  \href{http://dx.doi.org/10.21468/SciPostPhys.10.2.050}{{\em SciPost Phys.}
  {\bfseries 10} no.~2, (2021) 050},
  \href{http://arxiv.org/abs/2007.04990}{{\ttfamily arXiv:2007.04990
  [hep-ph]}}.

\bibitem{Dine:2020pds}
M.~Dine, N.~Fernandez, A.~Ghalsasi, and H.~H. Patel, ``{Comments on axions,
  domain walls, and cosmic strings},''
  \href{http://dx.doi.org/10.1088/1475-7516/2021/11/041}{{\em JCAP} {\bfseries
  11} (2021) 041}, \href{http://arxiv.org/abs/2012.13065}{{\ttfamily
  arXiv:2012.13065 [hep-ph]}}.

\bibitem{Buschmann:2021sdq}
M.~Buschmann, J.~W. Foster, A.~Hook, A.~Peterson, D.~E. Willcox, W.~Zhang, and
  B.~R. Safdi, ``{Dark matter from axion strings with adaptive mesh
  refinement},'' \href{http://dx.doi.org/10.1038/s41467-022-28669-y}{{\em
  Nature Commun.} {\bfseries 13} no.~1, (2022) 1049},
  \href{http://arxiv.org/abs/2108.05368}{{\ttfamily arXiv:2108.05368
  [hep-ph]}}.

\bibitem{Agrawal:2019lkr}
P.~Agrawal, A.~Hook, and J.~Huang, ``{A CMB Millikan experiment with cosmic
  axiverse strings},'' \href{http://dx.doi.org/10.1007/JHEP07(2020)138}{{\em
  JHEP} {\bfseries 07} (2020) 138},
  \href{http://arxiv.org/abs/1912.02823}{{\ttfamily arXiv:1912.02823
  [astro-ph.CO]}}.

\bibitem{Benabou:2023ghl}
J.~N. Benabou, M.~Buschmann, S.~Kumar, Y.~Park, and B.~R. Safdi, ``{Signatures
  of Primordial Energy Injection from Axion Strings},''
  \href{http://arxiv.org/abs/2308.01334}{{\ttfamily arXiv:2308.01334
  [hep-ph]}}.

\bibitem{Callan:1984sa}
C.~G. Callan, Jr. and J.~A. Harvey, ``{Anomalies and Fermion Zero Modes on
  Strings and Domain Walls},''
  \href{http://dx.doi.org/10.1016/0550-3213(85)90489-4}{{\em Nucl. Phys. B}
  {\bfseries 250} (1985) 427--436}.

\bibitem{Kaplan:1987kh}
D.~B. Kaplan and A.~Manohar, ``{Anomalous Vortices and Electromagnetism},''
  \href{http://dx.doi.org/10.1016/0550-3213(88)90244-1}{{\em Nucl. Phys. B}
  {\bfseries 302} (1988) 280--290}.

\bibitem{Naculich:1987ci}
S.~G. Naculich, ``{Axionic Strings: Covariant Anomalies and Bosonization of
  Chiral Zero Modes},''
  \href{http://dx.doi.org/10.1016/0550-3213(88)90400-2}{{\em Nucl. Phys. B}
  {\bfseries 296} (1988) 837--867}.

\bibitem{Manohar:1988gv}
A.~Manohar, ``{Anomalous Vortices and Electromagnetism. 2.},''
  \href{http://dx.doi.org/10.1016/0370-2693(88)91505-5}{{\em Phys. Lett. B}
  {\bfseries 206} (1988) 276}. [Erratum: Phys.Lett. 209, 543 (1988)].

\bibitem{Harvey:1988in}
J.~A. Harvey and S.~G. Naculich, ``{Cosmic Strings From Pseudoanomalous
  U(1)$s$},'' \href{http://dx.doi.org/10.1016/0370-2693(89)90857-5}{{\em Phys.
  Lett. B} {\bfseries 217} (1989) 231--237}.

\bibitem{Harari:1992ea}
D.~Harari and P.~Sikivie, ``{Effects of a Nambu-Goldstone boson on the
  polarization of radio galaxies and the cosmic microwave background},''
  \href{http://dx.doi.org/10.1016/0370-2693(92)91363-E}{{\em Phys. Lett. B}
  {\bfseries 289} (1992) 67--72}.

\bibitem{Blum:1993yd}
J.~D. Blum and J.~A. Harvey, ``{Anomaly inflow for gauge defects},''
  \href{http://dx.doi.org/10.1016/0550-3213(94)90580-0}{{\em Nucl. Phys. B}
  {\bfseries 416} (1994) 119--136},
  \href{http://arxiv.org/abs/hep-th/9310035}{{\ttfamily arXiv:hep-th/9310035}}.

\bibitem{Harvey:2000yg}
J.~A. Harvey and O.~Ruchayskiy, ``{The Local structure of anomaly inflow},''
  \href{http://dx.doi.org/10.1088/1126-6708/2001/06/044}{{\em JHEP} {\bfseries
  06} (2001) 044}, \href{http://arxiv.org/abs/hep-th/0007037}{{\ttfamily
  arXiv:hep-th/0007037}}.

\bibitem{Heidenreich:2021yda}
B.~Heidenreich, M.~Reece, and T.~Rudelius, ``{The Weak Gravity Conjecture and
  axion strings},'' \href{http://dx.doi.org/10.1007/JHEP11(2021)004}{{\em JHEP}
  {\bfseries 11} (2021) 004}, \href{http://arxiv.org/abs/2108.11383}{{\ttfamily
  arXiv:2108.11383 [hep-th]}}.

\bibitem{Fukuda:2020imw}
H.~Fukuda and K.~Yonekura, ``{Witten effect, anomaly inflow, and charge
  teleportation},'' \href{http://dx.doi.org/10.1007/JHEP01(2021)119}{{\em JHEP}
  {\bfseries 01} (2021) 119}, \href{http://arxiv.org/abs/2010.02221}{{\ttfamily
  arXiv:2010.02221 [hep-th]}}.

\bibitem{Fukuda:2020kym}
H.~Fukuda, A.~V. Manohar, H.~Murayama, and O.~Telem, ``{Axion strings are
  superconducting},'' \href{http://dx.doi.org/10.1007/JHEP06(2021)052}{{\em
  JHEP} {\bfseries 06} (2021) 052},
  \href{http://arxiv.org/abs/2010.02763}{{\ttfamily arXiv:2010.02763
  [hep-ph]}}.

\bibitem{Ibe:2021ctf}
M.~Ibe, S.~Kobayashi, Y.~Nakayama, and S.~Shirai, ``{On Stability of Fermionic
  Superconducting Current in Cosmic String},''
  \href{http://dx.doi.org/10.1007/JHEP05(2021)217}{{\em JHEP} {\bfseries 05}
  (2021) 217}, \href{http://arxiv.org/abs/2102.05412}{{\ttfamily
  arXiv:2102.05412 [hep-ph]}}.

\bibitem{Agrawal:2020euj}
P.~Agrawal, A.~Hook, J.~Huang, and G.~Marques-Tavares, ``{Axion string
  signatures: a cosmological plasma collider},''
  \href{http://dx.doi.org/10.1007/JHEP01(2022)103}{{\em JHEP} {\bfseries 01}
  (2022) 103}, \href{http://arxiv.org/abs/2010.15848}{{\ttfamily
  arXiv:2010.15848 [hep-ph]}}.

\bibitem{Witten:1984eb}
E.~Witten, ``{Superconducting Strings},''
  \href{http://dx.doi.org/10.1016/0550-3213(85)90022-7}{{\em Nucl. Phys. B}
  {\bfseries 249} (1985) 557--592}.

\bibitem{Dine:1981rt}
M.~Dine, W.~Fischler, and M.~Srednicki, ``{A Simple Solution to the Strong CP
  Problem with a Harmless Axion},''
  \href{http://dx.doi.org/10.1016/0370-2693(81)90590-6}{{\em Phys. Lett. B}
  {\bfseries 104} (1981) 199--202}.

\bibitem{Zhitnitsky:1980tq}
A.~R. Zhitnitsky, ``{On Possible Suppression of the Axion Hadron Interactions.
  (In Russian)},'' {\em Sov. J. Nucl. Phys.} {\bfseries 31} (1980) 260.

\bibitem{Abe:2020ure}
Y.~Abe, Y.~Hamada, and K.~Yoshioka, ``{Electroweak axion string and
  superconductivity},'' \href{http://dx.doi.org/10.1007/JHEP06(2021)172}{{\em
  JHEP} {\bfseries 06} (2021) 172},
  \href{http://arxiv.org/abs/2010.02834}{{\ttfamily arXiv:2010.02834
  [hep-ph]}}.

\bibitem{Hill:1986ts}
C.~T. Hill and L.~M. Widrow, ``{Superconducting Cosmic Strings with Massive
  Fermions},'' \href{http://dx.doi.org/10.1016/0370-2693(87)91262-7}{{\em Phys.
  Lett. B} {\bfseries 189} (1987) 17--22}.

\bibitem{Weldon:1982bn}
H.~A. Weldon, ``{Effective Fermion Masses of Order gT in High Temperature Gauge
  Theories with Exact Chiral Invariance},''
  \href{http://dx.doi.org/10.1103/PhysRevD.26.2789}{{\em Phys. Rev. D}
  {\bfseries 26} (1982) 2789}.

\bibitem{Achucarro:1999it}
A.~Achucarro and T.~Vachaspati, ``{Semilocal and electroweak strings},''
  \href{http://dx.doi.org/10.1016/S0370-1573(99)00103-9}{{\em Phys. Rept.}
  {\bfseries 327} (2000) 347--426},
  \href{http://arxiv.org/abs/hep-ph/9904229}{{\ttfamily arXiv:hep-ph/9904229}}.

\bibitem{Jackiw:1981ee}
R.~Jackiw and P.~Rossi, ``{Zero Modes of the Vortex - Fermion System},''
  \href{http://dx.doi.org/10.1016/0550-3213(81)90044-4}{{\em Nucl. Phys. B}
  {\bfseries 190} (1981) 681--691}.

\bibitem{Starkman:2000bq}
G.~D. Starkman, D.~Stojkovic, and T.~Vachaspati, ``{Neutrino zero modes on
  electroweak strings},''
  \href{http://dx.doi.org/10.1103/PhysRevD.63.085011}{{\em Phys. Rev. D}
  {\bfseries 63} (2001) 085011},
  \href{http://arxiv.org/abs/hep-ph/0007071}{{\ttfamily arXiv:hep-ph/0007071}}.

\bibitem{Stojkovic:2000ix}
D.~Stojkovic, ``{Neutrino zero modes and stability of electroweak strings},''
  \href{http://dx.doi.org/10.1142/S0217751X01008813}{{\em Int. J. Mod. Phys. A}
  {\bfseries 16S1C} (2001) 1034--1036},
  \href{http://arxiv.org/abs/hep-th/0103216}{{\ttfamily arXiv:hep-th/0103216}}.

\bibitem{Starkman:2001tc}
G.~Starkman, D.~Stojkovic, and T.~Vachaspati, ``{Zero modes of fermions with a
  general mass matrix},''
  \href{http://dx.doi.org/10.1103/PhysRevD.65.065003}{{\em Phys. Rev. D}
  {\bfseries 65} (2002) 065003},
  \href{http://arxiv.org/abs/hep-th/0103039}{{\ttfamily arXiv:hep-th/0103039}}.

\bibitem{Naculich:1995cb}
S.~G. Naculich, ``{Fermions destabilize electroweak strings},''
  \href{http://dx.doi.org/10.1103/PhysRevLett.75.998}{{\em Phys. Rev. Lett.}
  {\bfseries 75} (1995) 998--1001},
  \href{http://arxiv.org/abs/hep-ph/9501388}{{\ttfamily arXiv:hep-ph/9501388}}.

\bibitem{Liu:1995at}
H.~Liu and T.~Vachaspati, ``{Perturbed electroweak strings and fermion zero
  modes},'' \href{http://dx.doi.org/10.1016/0550-3213(96)00158-7}{{\em Nucl.
  Phys. B} {\bfseries 470} (1996) 176--194},
  \href{http://arxiv.org/abs/hep-ph/9511216}{{\ttfamily arXiv:hep-ph/9511216}}.

\bibitem{Coleman:1973jx}
S.~R. Coleman and E.~J. Weinberg, ``{Radiative Corrections as the Origin of
  Spontaneous Symmetry Breaking},''
  \href{http://dx.doi.org/10.1103/PhysRevD.7.1888}{{\em Phys. Rev. D}
  {\bfseries 7} (1973) 1888--1910}.

\bibitem{Agrawal:2023sbp}
P.~Agrawal and A.~Platschorre, ``{The Monodromic Axion-Photon Coupling},''
  \href{http://arxiv.org/abs/2309.03934}{{\ttfamily arXiv:2309.03934
  [hep-th]}}.

\bibitem{Bender:1999amm}
C.~Bender and S.~Orszag, {\em Advanced Mathematical Methods for Scientists and
  Engineers I: Asymptotic Methods and Perturbation Theory}.
\newblock Springer, 1999.

\bibitem{Kaplan:1992sg}
D.~B. Kaplan, ``{Chiral fermions on the lattice},''
  \href{http://dx.doi.org/10.1016/0920-5632(93)90282-B}{{\em Nucl. Phys. B
  Proc. Suppl.} {\bfseries 30} (1993) 597--600}.

\bibitem{Aitken:2017nfd}
K.~Aitken, A.~Baumgartner, A.~Karch, and B.~Robinson, ``{3d Abelian Dualities
  with Boundaries},'' \href{http://dx.doi.org/10.1007/JHEP03(2018)053}{{\em
  JHEP} {\bfseries 03} (2018) 053},
  \href{http://arxiv.org/abs/1712.02801}{{\ttfamily arXiv:1712.02801
  [hep-th]}}.

\bibitem{Fradkin:2013sab}
E.~H. Fradkin, {\em {Field Theories of Condensed Matter Physics}}, vol.~82.
\newblock Cambridge Univ. Press, Cambridge, UK, 2, 2013.

\bibitem{Haldane:1988zza}
F.~D.~M. Haldane, ``{Model for a Quantum Hall Effect without Landau Levels:
  Condensed-Matter Realization of the 'Parity Anomaly'},''
  \href{http://dx.doi.org/10.1103/PhysRevLett.61.2015}{{\em Phys. Rev. Lett.}
  {\bfseries 61} (1988) 2015--2018}.

\bibitem{Jansen:1992tw}
K.~Jansen and M.~Schmaltz, ``{Critical momenta of lattice chiral fermions},''
  \href{http://dx.doi.org/10.1016/0370-2693(92)91335-7}{{\em Phys. Lett. B}
  {\bfseries 296} (1992) 374--378},
  \href{http://arxiv.org/abs/hep-lat/9209002}{{\ttfamily
  arXiv:hep-lat/9209002}}.

\bibitem{Witten:2019bou}
E.~Witten and K.~Yonekura, ``{Anomaly Inflow and the $\eta$-Invariant},'' in
  {\em {The Shoucheng Zhang Memorial Workshop}}.
\newblock 9, 2019.
\newblock \href{http://arxiv.org/abs/1909.08775}{{\ttfamily arXiv:1909.08775
  [hep-th]}}.

\bibitem{Goldstone:1981kk}
J.~Goldstone and F.~Wilczek, ``{Fractional Quantum Numbers on Solitons},''
  \href{http://dx.doi.org/10.1103/PhysRevLett.47.986}{{\em Phys. Rev. Lett.}
  {\bfseries 47} (1981) 986--989}.

\bibitem{Ross:2012fc}
A.~Ross, ``{Multipole expansion at the level of the action},''
  \href{http://dx.doi.org/10.1103/PhysRevD.85.125033}{{\em Phys. Rev. D}
  {\bfseries 85} (2012) 125033},
  \href{http://arxiv.org/abs/1202.4750}{{\ttfamily arXiv:1202.4750 [gr-qc]}}.

\bibitem{Callan:1982ah}
C.~G. Callan, Jr., ``{Disappearing Dyons},''
  \href{http://dx.doi.org/10.1103/PhysRevD.25.2141}{{\em Phys. Rev. D}
  {\bfseries 25} (1982) 2141}.

\bibitem{Trefethen:2013ata}
L.~Trefethen, {\em Approximation Theory and Approximation Practice}, vol.~128.
\newblock Siam, 2013.

\bibitem{Berrut:2004bli}
J.-P. Berrut and L.~Trefethen, ``{Barycentric Lagrange Interpolation},'' {\em
  SIAM review} {\bfseries 46} (2004) 501--517.

\bibitem{Higham:2004num}
N.~Higham, ``{The Numerical Stability of Barycentric Lagrange Interpolation},''
  {\em IMA Journal of Numerical Analysis} {\bfseries 24} (2004) 547--556.

\bibitem{Baltensperger:2000imp}
R.~Baltensperger, ``{Improving the Accuracy of the Matrix Differentiation
  Method for Arbitrary Collocation Points},'' {\em Applied Numerical
  Mathematics} {\bfseries 33} (2000) 143--149.

\end{thebibliography}\endgroup

\end{document}